\documentclass{emulateapj}
\usepackage{color}

\begin{document}

\shortauthors{Luhman}
\shorttitle{Stellar Membership of Taurus}

\title{The Stellar Membership of the Taurus Star-forming
Region\altaffilmark{1}}

\author{
K. L. Luhman\altaffilmark{2,3}}

\altaffiltext{1}
{Based on observations performed with the {\it Gaia} mission, the Two Micron
All Sky Survey, and the United Kingdom Infrared Telescope Infrared Deep
Sky Survey.}

\altaffiltext{2}{Department of Astronomy and Astrophysics, The Pennsylvania
State University, University Park, PA 16802; kluhman@astro.psu.edu}

\altaffiltext{3}{Center for Exoplanets and Habitable Worlds, The
Pennsylvania State University, University Park, PA 16802, USA}

\begin{abstract}

The high-precision astrometry from the second data release
of the {\it Gaia} mission has made it possible to greatly improve
the census of members of nearby clusters and associations.
I have applied the {\it Gaia} data to the Taurus star-forming region,
refining the sample of known members and identifying candidates for
undiscovered members.
The resulting samples of members and candidates provide the best constraints
to date on the distribution of ages and the initial mass function (IMF) in
Taurus. Several studies over the last 30 years have proposed the existence
of a population of older stars ($\gtrsim10$~Myr) that is associated
with the Taurus clouds. The data from {\it Gaia} demonstrate
that such a population does not exist. Meanwhile, previous IMF estimates
for small fields surrounding the Taurus aggregates have exhibited a
surplus of K7--M0 stars (0.7--0.8~$M_\odot$) relative to star-forming
clusters like IC~348 and the Orion Nebula Cluster. 
However, that difference disappears when the new census of the
entire region is considered, which should be complete for 
spectral types earlier than M6--M7 at $A_J<1$. Thus, there is little variation
in the stellar IMF across the 3--4 orders of magnitude in stellar density
that are present in nearby star-forming regions.
Finally, I note that the proper motions of two previously known members,
KPNO~15 and 2MASS J04355209+2255039,
indicate that they may have been ejected from the same location within
the L1536 cloud $\sim7200$ years ago.

\end{abstract}


\section{Introduction}
\label{sec:intro}

The Taurus star-forming region has served as one of the primary laboratories
for investigating the process of star formation.
This is in part due to its proximity to the Sun 
\citep[$d\sim140$~pc,][references therein]{gal18} and the relatively large 
size of its stellar population \citep[$N\sim400$,][]{ken08}.
The importance of Taurus also stems from the unusually wide distribution of
its young stars such that a comparison of the long crossing time of the region
(10--20~Myr) to the age spread among its members places stringent constraints
on theories for the formation of molecular clouds \citep{bal99,har01b}.
Meanwhile, given its low stellar density (1--10~pc$^{-3}$), Taurus can be used
to search for a variation of the initial mass function (IMF) with star-forming
conditions \citep{bon11}.

Measuring the distributions of ages and masses in Taurus requires a thorough
census of its stellar population.
Previous studies have searched for members of Taurus using signatures of youth
(variability, emission lines, infrared (IR) excess emission, X-ray emission),
proper motions, and optical and near-IR color-magnitude diagrams 
\citep[][references therein]{luh17}.
Those surveys have tended to be less sensitive to stars at older ages, but
they have demonstrated that Taurus is unlikely to contain a large population
of stars with ages of $\gtrsim10$~Myr \citep{har91,gom92}.
Modest numbers of stars with ages from $\sim$10--100~Myr have been found in the
direction of Taurus, ranging from early-type stars \citep{bla56} to brown
dwarfs \citep{luh06tau1,sle06,esp17}, some of which have been proposed to
be products of the Taurus clouds \citep{wal88,neu95,wic96,dae15,kra17,zha18}.
However, the surface densities and ages of those stars found in X-ray surveys
are consistent with members of the solar neighborhood \citep{bri97} while most
of the stars with precise proper motions are kinematically distinct from
the younger stars associated with the clouds \citep{har91,fri97,dez99,esp17}.
Nevertheless, well-defined constraints on the size of an older population in 
Taurus are not yet available.

Surveys for members of Taurus also have been used to derive estimates of the
stellar IMF in the region \citep{bri02,luh03tau,luh04tau,luh09tau}.
Those studies have found that the richest stellar aggregates exhibit a 
surplus of K7--M0 stars (0.7--0.8~$M_\odot$) relative to clusters with higher 
stellar densities like the Orion Nebula Cluster (ONC) and IC~348
\citep{hil97,hc00,mue02,mue03,luh03ic}.
That surplus appears to be somewhat less pronounced when a larger area of 
Taurus is considered \citep{luh17}.
The most definitive comparison to other star-forming regions would employ the 
IMF for the entire cloud complex in Taurus, but a reliable measurement 
has not been possible because of uncertainties in the completeness of the
current census.

As with a multitude of other topics, the astrometry from the
{\it Gaia} mission \citep{per01,deb12} offers an opportunity for
dramatic progress on obtaining a complete census of Taurus. 
The second data release of {\it Gaia} (DR2) contains an all-sky catalog
of parallaxes and proper motions with errors of $\lesssim0.7$~mas and
$\lesssim1.2$~mas~yr$^{-1}$, respectively, for stars at $G\lesssim20$
\citep{bro18}, which corresponds to errors of $\lesssim10$\% and $\lesssim5$\%
for members of Taurus at $\gtrsim0.05$~$M_\odot$ that have low extinction.
Thus, the {\it Gaia} data can be used for precise kinematic identification
of undiscovered members of Taurus across the entire range of stellar masses.
Since {\it Gaia} operates at optical wavelengths, heavily reddened members
can fall below its detection limit, but it is the areas of high extinction
near the clouds have been most thoroughly searched for members in previous
surveys, so most of the missing members (particularly older ones) are likely
outside of the clouds where the extinction is low.
In this paper, I have compiled the {\it Gaia} parallaxes and proper motions
for the known members of Taurus adopted by \citet{esp17}, characterized the
kinematics and distances of those objects, and checked that catalog for
nonmembers (Section~\ref{sec:known}).
The {\it Gaia} data are used to assess the membership of previous samples of 
older stars in the direction of Taurus (Section~\ref{sec:oldsamples}) and 
search for new members at any age
(Section~\ref{sec:search}). Using the refined census of known members
and the new candidates from {\it Gaia}, I estimate the distribution of
ages and the IMF for Taurus (Section~\ref{sec:stellarpop}).

\section{Kinematics of Known Taurus Members}
\label{sec:known}

\subsection{Retrieval of Data from {\it Gaia} DR2}

All stars in DR2 from {\it Gaia} have single-epoch positions and 
photometry in a broad optical band ($G$, 3300--10500~\AA). Most of those
stars also have data in bands at 3300--6800 and 6300-10500~\AA\ ($G_{\rm BP}$ 
and $G_{\rm RP}$). Proper motions and parallaxes are available for most stars
down to $G\sim20$ and radial velocities are available primarily for stars
at $G\sim4$--12. Additional data products from DR2 are described by 
\citet{bro18}.

To examine the kinematics and distances of known members of Taurus,
I have considered the 427 stars that were adopted as members by \citet{esp17}.
For each star, I identified the closest counterpart in {\it Gaia} DR2 
within $1\arcsec$ of its position in the 2MASS Point Source Catalog.
Some members had multiple {\it Gaia} counterparts, all of which corresponded
to the known components of binary systems.
I retrieved the photometry, parallaxes, proper motions, and radial
velocities for those counterparts from DR2. Multiple systems in
which the components were unresolved from each other in the photometric
catalogs utilized by \citet{esp17} appeared as single entries in the list of
members from that study. {\it Gaia} provides resolved measurements for
companions in 13 of those systems, so they are now counted as separate objects.
They consist of FQ~Tau~B, UX~Tau~C, FX~Tau~B, IRAS~04278+2253~B, GG~Tau~Bb,
HN~Tau~B, CoKu~Tau~3~B, GN~Tau~B, CIDA 9~B, RW~Aur~B, HBC~358~B, GZ~Aur~B,
and BS~Tau~B.

Among the 440 objects from \citet{esp17}, 382, 336, and 28 stars have
positions, parallaxes/proper motions, and radial velocities from DR2,
respectively. One source, GG~Tau~Aa+Ab, has a negative parallax, which is 
likely due to its binarity. Its parallax and proper motion are excluded
from this work. The median errors in parallax,
proper motion, and radial velocity for this sample are $\sim$0.1~mas,
0.2~mas~yr$^{-1}$, and 4~km~s$^{-1}$, respectively.
For most of these stars, the errors in parallax and proper motion are much
smaller than those from previous measurements.
Prior to {\it Gaia}, the most accurate parallaxes and proper motions in
Taurus were measured with the Very Long Baseline Interferometry (VLBI),
which produced errors comparable to that from {\it Gaia} for 16 
systems\footnote{Those systems consist of
Anon~1 (V1096~Tau), V773~Tau, LkCa~3 (V1098~Tau), V410 Anon 25,
V410 Tau (HD~283518), Hubble~4 (V1023~Tau), T~Tau, RX J0424.8+2643
(V1201~Tau), HD~283641, XZ~Tau, V807~Tau, HP~Tau~G2, 
LkHa~332/G1 (V1000~Tau), LkHa~332/G2 (V999~Tau), LkCa~19 (HD~282630),
and HD~283572.} \citep{loi05,loi07,tor07,tor09,tor12,gal18}.
The {\it Gaia} radial velocities in Taurus have limited value
given that ground-based studies have measured more accurate velocities
for a larger number of members \citep[e.g.,][]{har86,whi03,ngu12}.

\subsection{Analysis of {\it Gaia} Data}
\label{sec:pops}

\begin{figure}
\epsscale{1}
\plotone{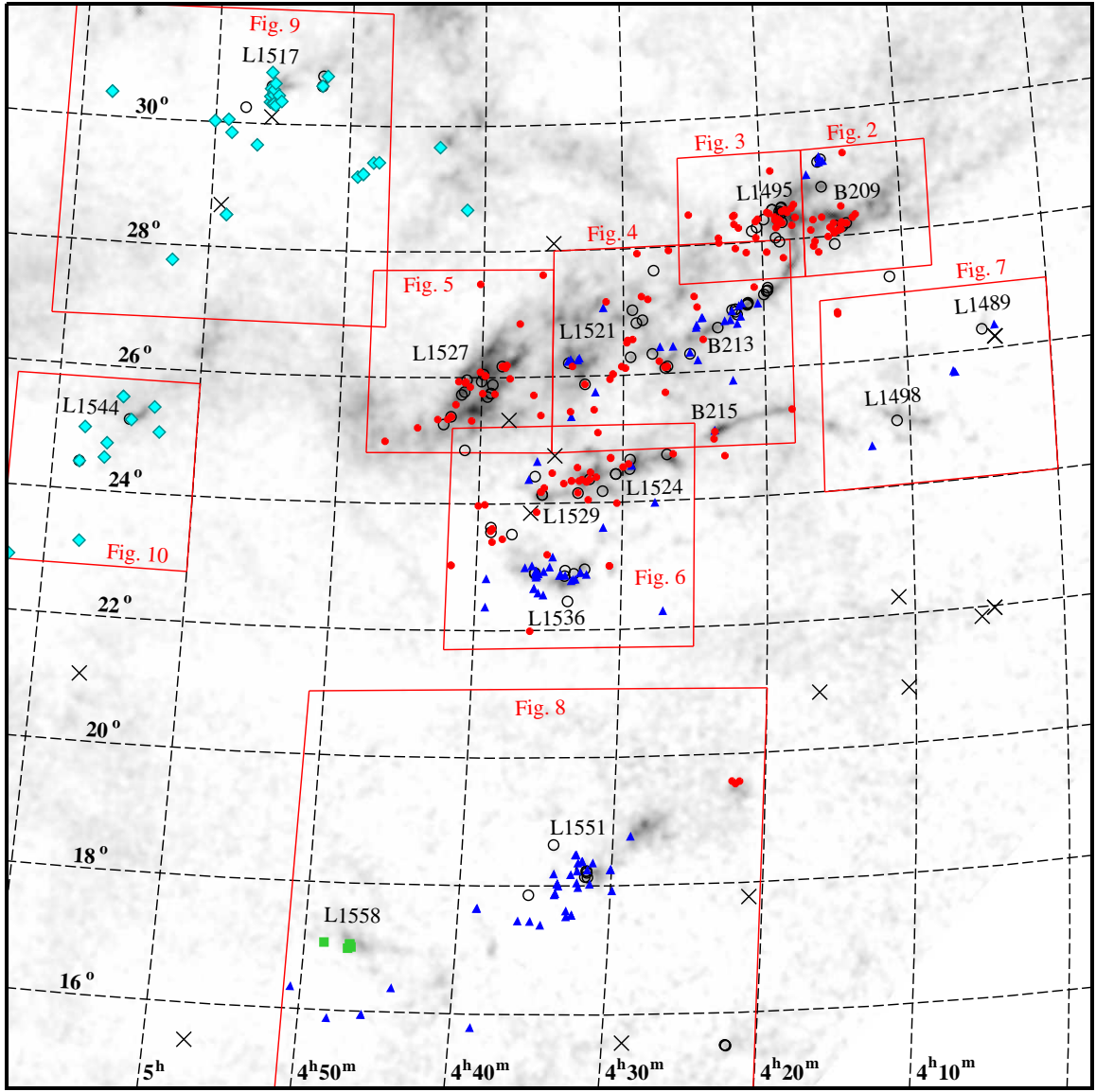}
\caption{
Spatial distribution of stars adopted as members of Taurus by \citet{esp17}.
Stars that have parallaxes and proper motions from {\it Gaia} DR2 are shown
with filled symbols (red circles, blue triangles, green squares, cyan diamonds)
if they are retained as members in this work or crosses if they are rejected 
as nonmembers. Members that lack {\it Gaia} data are plotted with open circles.
The filled symbols are assigned based on the kinematic populations in
Figures~\ref{fig:map1}--\ref{fig:map15}. The boundaries of the fields
encompassed by those figures are marked by the red rectangles.
The dark clouds in Taurus are displayed with a map of extinction
\citep[gray scale,][]{dob05}.  Designations for some of the most
prominent clouds are indicated \citep{bar27,lyn62}.
}
\label{fig:map}
\end{figure}

The positions of the members of Taurus from \citet{esp17} are plotted
with a map of extinction in Figure~\ref{fig:map}.
The area covered by these stars has a diameter of more than $10\arcdeg$,
which corresponds to $>$25~pc at their distances. One would expect a comparable 
spread in line-of-sight distances among the members. Indeed, \citet{gal18}
has recently detected such a spread using parallaxes for 26 systems from 
VLBI and the first data release of {\it Gaia} (DR1).
The kinematics of the stars may also vary noticeably across such a large
region. Therefore, to characterize the {\it Gaia} proper motions and
parallaxes of the known members, I have examined separately the stars
within nine fields that were selected to cover subsections of
the cloud complex.
The boundaries for the fields are indicated in Figure~\ref{fig:map}.

Data for the members within the nine fields are shown
in Figures~\ref{fig:map1}--\ref{fig:map11}. The few stars outside
of those fields are plotted in Figures~\ref{fig:map15}.
Each figure contains four diagrams for the members within a given field:
a map of the positions, extinction-corrected $M_J$ versus spectral type,
$G$ versus parallax, and proper motion offsets relative to the values
expected for the positions and parallaxes of the stars assuming the median
space velocity of Taurus members (Section~\ref{sec:revised}).
The latter three diagrams show only stars that have {\it Gaia} parallaxes and 
proper motions. Those stars have been divided into four populations of members
(labeled with red, blue, green, and cyan symbols) that exhibit distinct 
combinations of parallax and proper motion offsets and objects that are 
probable nonmembers based on these data (crosses), as discussed later in this 
section. The diagrams of $M_J$ versus spectral type have been constructed with
the extinctions, photometry, and spectral types adopted by \citet{luh17}
and \citet{esp17}. To facilitate comparison of the sequences in
$M_J$ versus spectral type among the fields, I have included the median
sequences for Taurus and the Upper Sco association 
\citep[11~Myr,][]{pec12,fei16}. 
The sequence for the latter is based on the members compiled by \citet{luh18}.
Those diagrams of $M_J$ versus spectral type exclude stars identified later
in this section as having discrepant parallaxes.
For stars that appear to have erroneous parallaxes because of binarity or
that have parallax errors of $>10$\%, I have adopted the median parallax
of the kinematic population in their field to which they likely belong when
computing the proper motion offsets. The plotted errors in the offsets include
both the errors in the proper motions and the errors in the expected
motions due to the parallax measurements.
The expected motions are based on the median space velocity of Taurus
members that is derived in Section~\ref{sec:revised}.

I now discuss the four diagrams of data for each of the nine fields in Taurus.

\subsubsection{B209 (Figure~\ref{fig:map1})}

In the field containing the B209 cloud, the members form two distinct groups
in parallax with median values of 7.6 and 6.3~mas (132 and 158~pc).
These stars are plotted with red and blue symbols, respectively.
The two groups also differ in their locations and proper motion offsets.
The red population is projected against the B209 cloud and the smaller blue
group is $\sim1\arcdeg$ north of the cloud. 
The parallax errors for the three faintest stars in this field are too large
for identification of their respective groups, so I have assigned them
to the red population based on their locations and proper motion offsets.

A few members in this field have proper motions or parallaxes that differ
significantly from those of the groups. 
MHO~3 is discrepant relative to both groups in these parameters.
FO~Tau is located near the stars in the red group but its parallax agrees
better with that of the blue group.
The parallax and location of 2MASS J04163911+2858491 are indicative of the
blue population, but it has a discrepant proper motion relative to those stars.
These three stars exhibit unusually poor astrometric fits among Taurus members
based on the high values of the {\it Gaia} DR2 parameters 
{\tt astrometric\_gof\_al} and {\tt astrometric\_excess\_noise}, which can
be caused by the presence of a poorly resolved binary.
Indeed, FO~Tau and MHO~3 are known to be be tight binaries \citep{whi01,kra11b}.
Therefore, I have ignored the parallaxes of these three stars, continued to
treat them as members of Taurus, and assigned them to populations based on
their locations.

\subsubsection{L1495 (Figure~\ref{fig:map2})}

Most of the members in the field for L1495 comprise a single
group in both parallax and proper motion offset. 
They resemble the red population from B209 in both parameters and have
a median parallax of 7.8~mas (128~pc).
Two stars, RY~Tau and IRAS~04158+2805, have discrepant parallaxes.
The astrometric fits for these stars are poor according to 
{\tt astrometric\_gof\_al} and {\tt astrometric\_excess\_noise},
so both stars are retained as members and included in the red population.
The poor fits may indicate the presence of tight binaries. \citet{ngu12}
identified RY~Tau as a possible spectroscopic binary. IRAS~04158+2805 exhibits
extended emission, which also may have affected the astrometry.

\subsubsection{L1521, B213, and B215 (Figure~\ref{fig:map8})}

The stars in the field encompassing L1521, B213, and B215 are clustered in 
parallax and proper motion offset like the two populations in B209 and L1495. 
The B213 and B215 filaments contain concentrations
of stars from the blue and red groups, respectively, while the remaining
members of the groups are intermingled in a wider distribution.
The red and blue stars have median parallaxes of 7.6 and 6.2~mas
(131 and 161~pc), respectively.
As in B209, the faintest stars have uncertain parallaxes that are consistent
with both populations, so they have been assigned to groups based on their
proper motion offsets.

J1-4872~A appears outside of the boundaries of the diagram of proper motion
offsets in Figure~\ref{fig:map8}. It has a moderately poor astrometric fit, so
that measurement could be erroneous. Its companion, which has a separation
of $3\farcs4$, lacks a proper motion measurement from {\it Gaia}.
I have retained the two stars as members.
DF~Tau has a moderately discrepant proper motion offset, which is likely
due to its very poor astrometric fit.

\subsubsection{L1527 (Figure~\ref{fig:map3})}

The stars in L1527 have a single moderately broad distribution of parallaxes.
The latter exhibit a gradient with right ascension, varying from $\sim8$
to 7~mas between the western and eastern boundaries of the field, which
indicates that the stars are related to the red population in
the adjacent field to the west that contains L1527. 
That relationship is supported by the similar proper motion offsets.
As a result, the stars in this field have been assigned to the red population.

The parallax of 2MASS J04380191+2519266, labeled with a cross in 
Figure~\ref{fig:map3}, is much smaller than that of other members. 
It was identified as a candidate
member of Taurus based on mid-IR excess emission \citep{reb10} and
spectroscopically classified as late K or early M \citep{reb10,esp14}.
The IR excess has served as the only evidence of its youth, and hence its
membership. A second object has been detected at $1\arcsec$ from 
2MASS J04380191+2519266 by {\it Gaia}, Pan-STARRS1 \citep{kai02,kai10},
and the United Kingdom Infrared Telescope Infrared Deep Sky Survey
\citep[UKIDSS,][]{law07}. It is fainter by $\sim2$~mag in those data and is
unresolved in the images from the {\it Spitzer Space Telescope}
that exhibited the mid-IR excess emission.  As discussed earlier, close 
pairs of objects can have erroneous astrometry. However, the {\it Gaia} DR2
parameters {\tt astrometric\_gof\_al} and {\tt astrometric\_excess\_noise} 
indicate a good astrometric fit, so the parallax should be reliable.
Therefore, it seems likely that 2MASS J04380191+2519266 is a field
star and the mid-IR excess arises from its neighbor (perhaps a red galaxy).

V955~Tau, 2MASS J04401447+2729112, and 2MASS J04354526+2737130 have
discrepant proper motion offsets, as shown in Figure~\ref{fig:map3}.
V955~Tau is a close binary \citep{lei93} and has a poor astrometric fit,
so its proper motion is probably unreliable.
The other two stars are not known binaries and have astrometric fits that
are comparable to those of most other members. They are the northernmost
stars in this field, and thus are farthest from other members.
They could be nonmembers, but given their fairly large astrometric errors,
they are retained as members for this study.

\subsubsection{L1524, L1529, and L1536 (Figure~\ref{fig:map4})}

The field with L1524, L1529, and L1536 has two populations that are similar
to the red and blue groups from previously discussed clouds.
Most of the red and blue stars in this field are projected against
L1524/L1529 and L1536, respectively. They have median parallaxes of
7.8 and 6.2~mas (128 and 161~pc).

2MASS J04362151+2351165 and 2MASS J04344586+2445145 (crosses in
Figure~\ref{fig:map4}) differ from other members in their parallaxes and
proper motion offsets. They are not known binaries and do not have
unusually poor astrometric fits relative to other Taurus members, so there 
is no basis for disregarding those measurements.
In addition, although the IR spectrum of 2MASS J04344586+2445145 
from \citet{luh17} was better matched by a young star than a field dwarf,
it lacks the Li absorption expected for the former in a spectrum from the
Large Sky Area Multi-Object Fiber Spectroscopic Telescope \citep{cui12,zha12}.
Both stars are treated as nonmembers in this work.

GV~Tau, KPNO~15, 2MASS J04355209+2255039, and HP~Tau/G2 are additional
outliers in the diagram of proper motion offsets in Figure~\ref{fig:map4}.
The measurement for GV~Tau is probably not reliable since it is a binary and
it has a poor astrometric fit. The other three stars have better fits
and are not known to be close binaries. All three of them reside
within one of the most compact groups of stars in this field.
The proper motion offsets of KPNO~15 and 2MASS J04355209+2255039
have roughly opposite directions. In fact, those data indicate that
the stars were near the same location $\sim$7200 years ago, as illustrated
in Figure~\ref{fig:ejected}. Thus, KPNO~15 and 2MASS J04355209+2255039
may have been participants in a dynamical interaction with one or more
additional stars \citep{pov67} that resulted in their ejection. 
An explanation for the discrepant motion of HP~Tau/G2 is less obvious, 
but the VLBI motion from \citet{gal18} agrees better with the group,
so it is retained as a member.

\subsubsection{L1489 and L1498 (Figure~\ref{fig:map12})}

Each of the small clouds L1489 and L1498 has a single star from \citet{esp17} 
projected against it. Both stars, IRAS 04016+2610 and 2MASS 04105425+2501266,
are highly reddened, so they lack {\it Gaia} data.
Several additional young stars are scattered more widely across this field,
which is west of the main complex of clouds in Taurus.
Two and four of those stars have similar parallaxes and proper motions as
the red and blue populations from the main cloud complex, respectively.
The two remaining systems in this field are HBC~358~ABC and HBC~359,
which have a separation of $20\arcsec$. HBC~358~A and BC are separated by
$1\farcs6$ and HBC~358~B and C are separated by $0\farcs15$ \citep{hk03}.
Parallaxes and proper motions from {\it Gaia} are available for 
HBC~358~BC and HBC~359, which are labeled with crosses in 
Figure~\ref{fig:map12}\footnote{Near-IR photometry is not available for
HBC~358~BC, so it does not appear in the diagram of $M_J$ versus spectral
type in Figure~\ref{fig:map12}.}. 
Their proper motion offsets differ from those of the populations
associated with Taurus clouds, so all members of these systems are
classified as nonmembers.

\subsubsection{L1551 and L1558 (Figure~\ref{fig:map13})}
\label{sec:map13}

The southernmost clouds in Taurus include L1558, L1551, the small cloud
near T~Tau, and cloud 18 from \citet{oni02}, which contains the protostars
IRAS~04191+1523 and IRAM~04191+1522. 
Most of the young stars in the field encompassing these clouds
form a well-defined group that resembles the blue populations from the
northern clouds except with slightly larger parallaxes and smaller
proper motion offsets in declination. The median parallax for that group
is 6.9~mas (145~pc). 
The two stars near T~Tau are similar to that population in parallax
but differ in their proper motion offsets. The latter are closer to the
values of the red populations in the previous fields, so the same
color is assigned to them. In proper motion, T~Tau agrees better with the
blue group than its two neighbors, so it is unclear to which population it
belongs. I have assigned it to the same group as its neighbors.
The five stars near L1558 have parallaxes and proper motion offsets that
are distinct from those of the red and blue groups, so they are labeled
with a third color of green. They have a median parallax of 5.1~mas (196~pc),
making them the most distant members of Taurus.

Some of the young stars in this field are outliers in parallax and proper
motion relative to the red, blue, and green groups.
HD~30171 is similar to the blue stars in terms of its proper motion offset
and is located $13\arcsec$ from a member of the blue group, IRAS 04429+1550.
Its parallax from DR2 is too small for that group (5.41$\pm$0.11~mas), but
its measurement from DR1 is in better agreement (7.07$\pm$0.24~mas).
Therefore, I have assigned it to the blue population.
A second star with a discrepant parallax is LkHa~358. It has a moderately
poor astrometric fit relative to other members of Taurus, which is
probably due to the presence of extended emission surrounding this protostar. 
I have ignored the {\it Gaia} measurement of its parallax and have retained
it as a member.
Haro~6-37~A is one of the stars near L1558 labeled in green. Its parallax
differs from that of the other green stars, but the parallax of its companion
Haro~6-37~B does agree with that group, so both stars are considered
members of it. Finally, J2-157 and 2MASS J04284199+1533535
(crosses in Figure~\ref{fig:map13})
differ from the groups in this field in both their parallaxes and proper motion 
offsets. The quality of the astrometric fit for each star is comparable to that 
of the fits for most Taurus members, so those measurements should be reliable.
They are excluded from my catalog of members.

\subsubsection{L1517 (Figure~\ref{fig:map10})}

Most of the young stars in the field for L1517 are members of a single
group in parallax and proper motion offset. The median parallax for
this group is 6.3~mas (159~pc). Some stars are clustered around L1517
while others are more widely distributed.
The distributions of parallax and proper motion offsets overlap with those
of the red and blue groups from other clouds, but they are sufficiently
distinct that I have labeled them with a fourth color of cyan.

2MASS J04485789+2913548, Haro~6-39, 2MASS J04555288+3006523,
and 2MASS J04591661+2840468 do not match the population in this field
in terms of either parallax or proper motion offset.
The first two stars have poor astrometric fits and the first is
$6\arcsec$ from a member of the cyan group (2MASS J04485745+2913521), so
I ignore their astrometry and treat them as members of that group.
2MASS J04555288+3006523 and 2MASS J04591661+2840468 have better astrometric
fits, so they are likely to be nonmembers. 
They are labeled with crosses in Figure~\ref{fig:map10}. 
2MASS J04555288+3006523 is beyond the boundaries of
the diagrams of parallax and proper motion offset.
2MASS J04591661+2840468 does not appear in the diagram of $M_J$ versus
spectral type since it lacks a spectral classification.

\subsubsection{L1544 (Figure~\ref{fig:map11})}

A single group in parallax and proper motion offset is present among the
stars in the L1544 field. It overlaps with the cyan population in L1517
in those parameters, so it has been assigned that color.
The median parallax is 5.8~mas (172~pc).

2MASS J05122759+2253492 and CIDA~11 are discrepant in their parallax and
proper motion, respectively. Both stars have poor astrometric fits, so they
are retained as members. They have been resolved as close pairs by {\it Gaia}
and \citet{kra11b}, respectively, which would explain the poor fits.
2MASS J05023985+2459337 also does not match the proper motion offset of
the group in this field. It does agree better with the motions of
the southernmost stars in the field for L1517, so it could be
a member of that population. The astrometric fit appears to be reliable.
It is unclear whether this star should be treated as a member of Taurus,
but I do so for the purposes of this work.

\subsubsection{Stars Outside of Previous Fields (Figure~\ref{fig:map15})}

Twelve stars from the catalog of members adopted by \citet{esp17} are
outside of the fields in Figures~\ref{fig:map1}--\ref{fig:map11}.
All but one have measurements of parallaxes and proper motions from
{\it Gaia} DR2. Eight of those 11 stars differ significantly
from the populations associated with the Taurus clouds in terms of their
parallaxes and proper motion offsets (crosses in Figure~\ref{fig:map15}).
They consist of HBC~360, HBC~361, HBC~362, 2MASS J04102834+2051507,
2MASS J04110570+2216313, 2MASS J04162725+20530, 2MASS J04345973+2807017,
and 2MASS J05064662+2104296.
The astrometric fits for the HBC~360 and HBC~361 are moderately poor while
the other stars have better fits. Those two stars have a separation of
$7\arcsec$ and are $14\arcmin$ from HBC~362. The three stars share similar
proper motions and parallaxes, which suggests that those measurements
are reliable (and that the stars are associated with each other).
These eight stars are rejected from my catalog of members. 

The remaining three stars with {\it Gaia} data are 2MASS J04225416+2439538,
CoKu~Tau~4, and CIDA~14. The astrometry for 2MASS J04225416+2439538 
agrees well with that of the red populations in the clouds that are
closest to it. CoKu~Tau~4 is a known binary with a separation of $0\farcs05$
\citep{ire08}, but its astrometric fit is not especially poor, so
its astrometry is probably reliable. Both CoKu~Tau~4 and CIDA~14 
are north of L1527 and southwest of L1517 and they have similar parallaxes
and proper motion offsets. In those parameters, the two stars are near
the clusters of values for the red and cyan populations in those clouds,
although they do not overlap with either group in both parameters
simultaneously. They agree slightly better with the cyan population in
L1517, so they have been assigned that color for the purposes of the figures.

HD~286178 is the one star from \citet{esp17} outside of the fields in 
Figures~\ref{fig:map1}--\ref{fig:map11} that lacks a parallax and proper
motion from {\it Gaia}. Those parameters were not measured because the
astrometric fit was very poor. Given its remote location relative to the
Taurus clouds and the presence of young stars across Taurus that are
kinematically unrelated to the clouds (Sections~\ref{sec:intro} and
\ref{sec:oldsamples}), it seems likely that HD~286178 is a nonmember,
so I treat it as such.

\subsection{Revised Catalog of Members}
\label{sec:revised}

As discussed in the previous section, I have rejected 19 of the
440 stars adopted as Taurus members by \citet{esp17}.
For reasons described in Sections~\ref{sec:oldsamples} and \ref{sec:search}, 
I also have assigned membership to 17 additional stars,
consisting of HD~28354, HD~283641, HD~283782, HD~30378, RX~J0422.1+1934,
L1551-55, RX~J0507.2+2437, JH~223~B, XEST~20-071~B, V892~Tau~NE,
2MASS J04284263+2714039~B, 2MASS J05080816+2427150~B,
PSO J065.8871+19.8386, PSO J071.6033+17.0281, PSO J071.3189+31.6888,
PSO J074.1999+29.2197, and PSO J076.2495+31.7503.
The revised catalog of 438 Taurus members is presented in Table~\ref{tab:mem}.

\begin{figure}
\epsscale{1.2}
\plotone{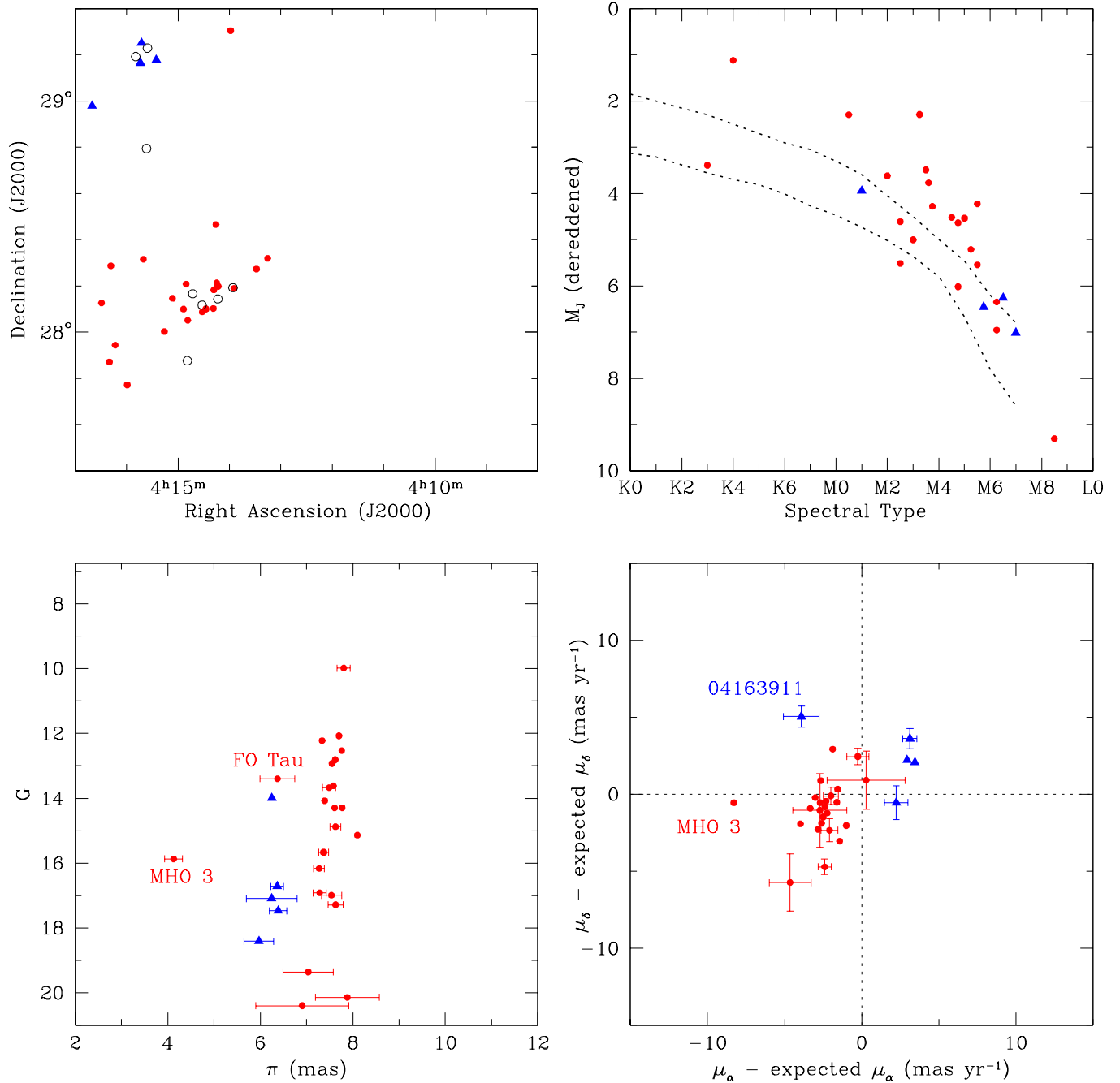}
\caption{
Spatial distribution of stars adopted as members of Taurus by \citet{esp17}
for a field encompassing the B209 cloud (top left). 
The symbols are the same as in Figure~\ref{fig:map}.
The stars that have parallaxes and proper motions from {\it Gaia} DR2
are shown in diagrams of extinction-corrected $M_J$
versus spectral type (top right), $G$ versus parallax (bottom left),
and proper motion offsets relative to the values expected for
the positions and parallaxes of the stars assuming the median space velocity
of Taurus members (bottom right). The diagram of $M_J$ versus spectral type
includes the median sequences for Taurus and Upper Sco (upper and lower
dotted lines).
Two stars in B209 with discrepant parallaxes and motions are labeled
in the bottom diagrams. They are omitted from $M_J$ versus spectral type.
Error bars are omitted in the bottom diagrams when the errors are smaller
than the symbols ($<0.1$~mas, $<0.5$~mas~yr$^{-1}$).
}
\label{fig:map1}
\end{figure}

\begin{figure}
\epsscale{1.2}
\plotone{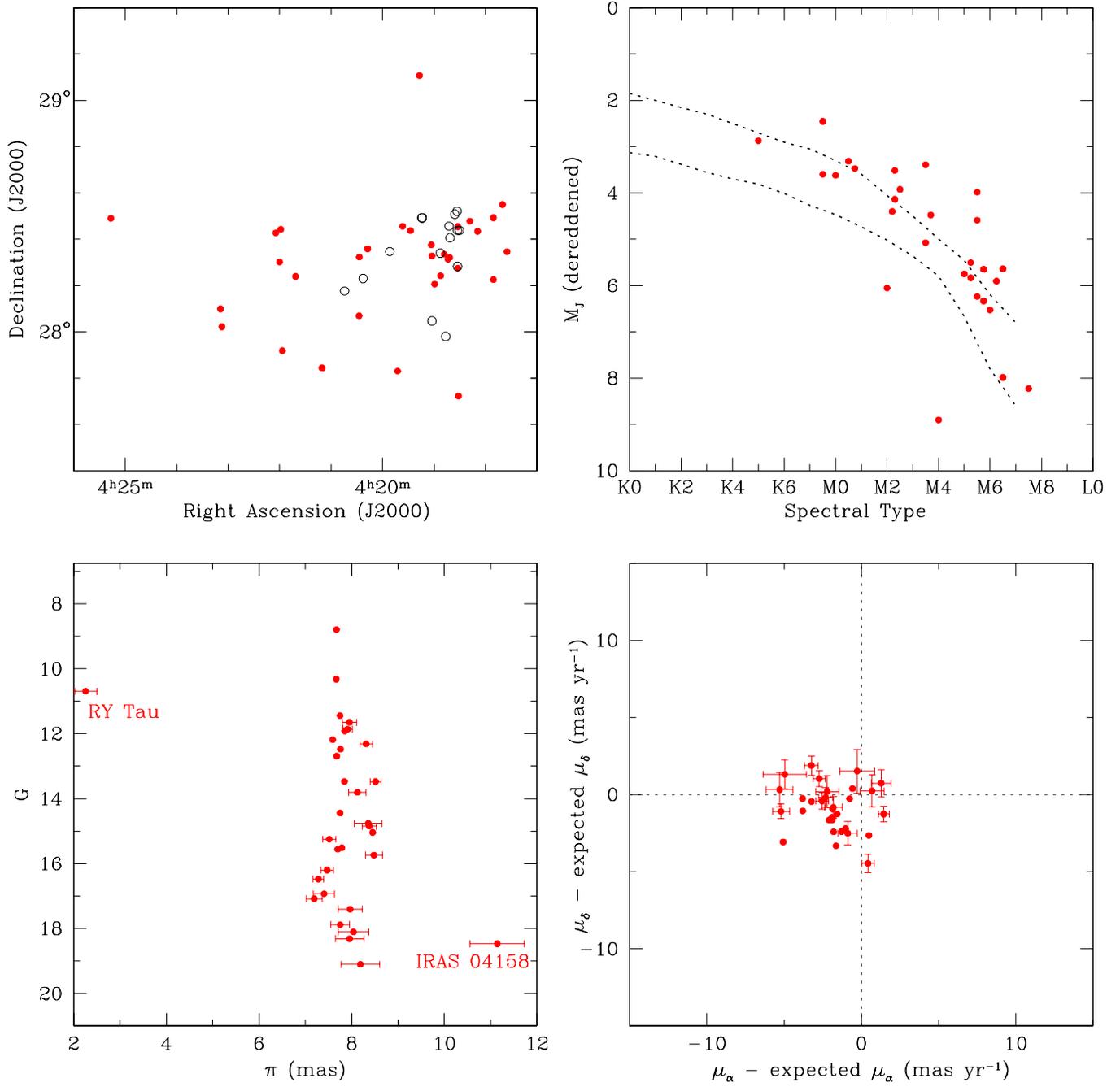}
\caption{
Same as Figure~\ref{fig:map1} for the L1495 cloud.
}
\label{fig:map2}
\end{figure}

\begin{figure}
\epsscale{1.2}
\plotone{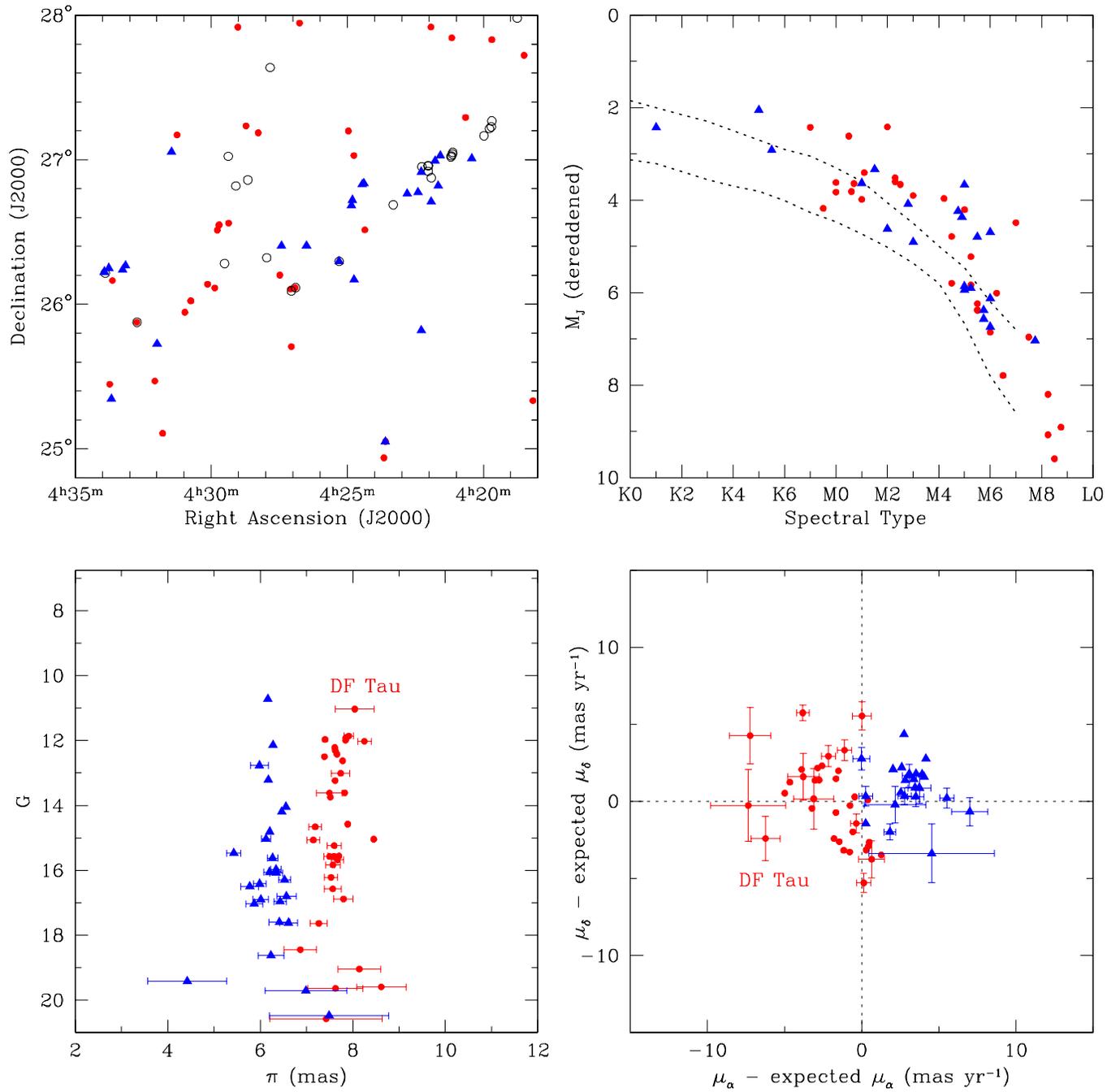}
\caption{
Same as Figure~\ref{fig:map1} for the L1521, B213, and B215 clouds.
}
\label{fig:map8}
\end{figure}

\begin{figure}
\epsscale{1.2}
\plotone{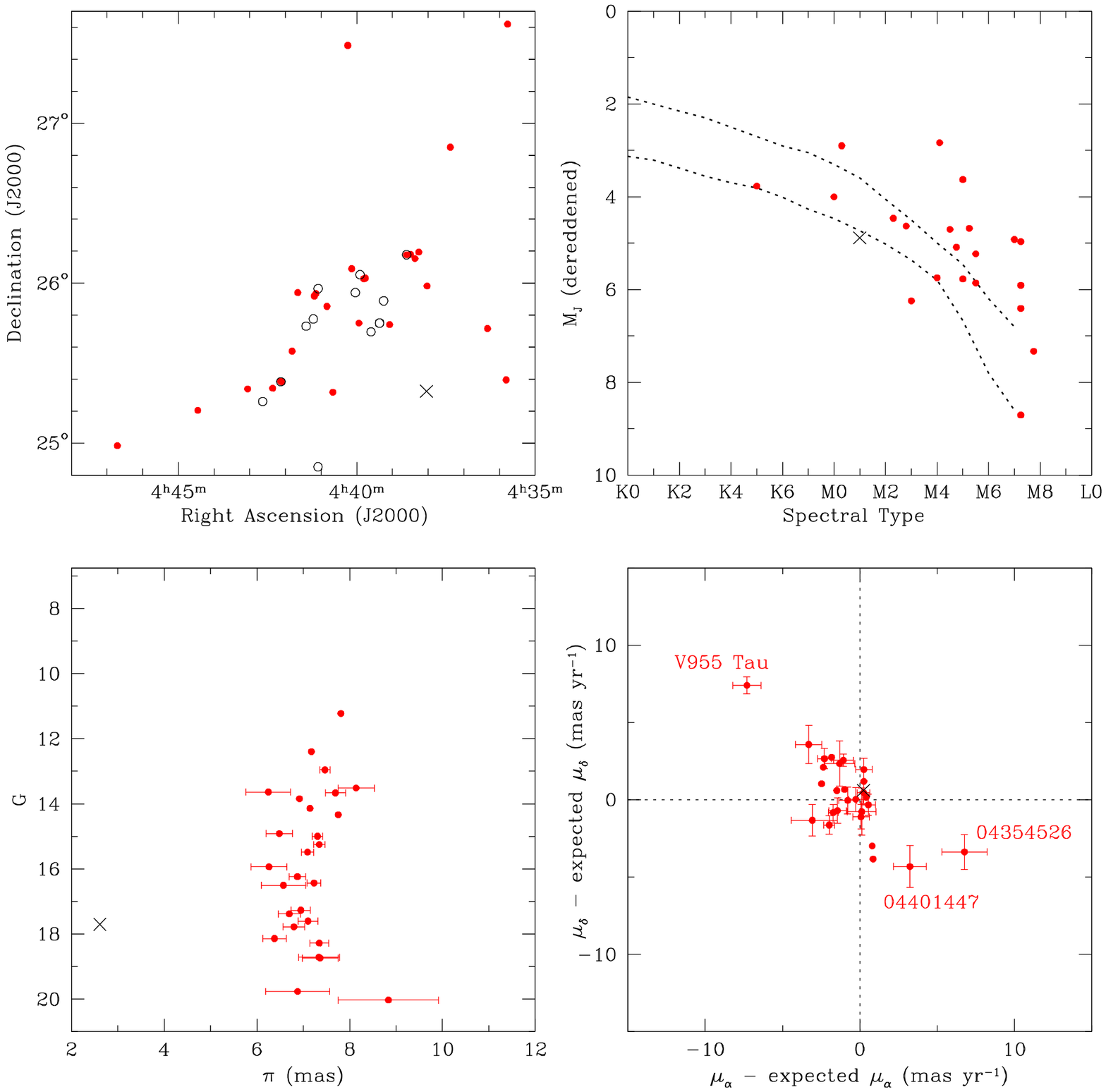}
\caption{
Same as Figure~\ref{fig:map1} for the L1527 cloud.
}
\label{fig:map3}
\end{figure}

\begin{figure}
\epsscale{1.2}
\plotone{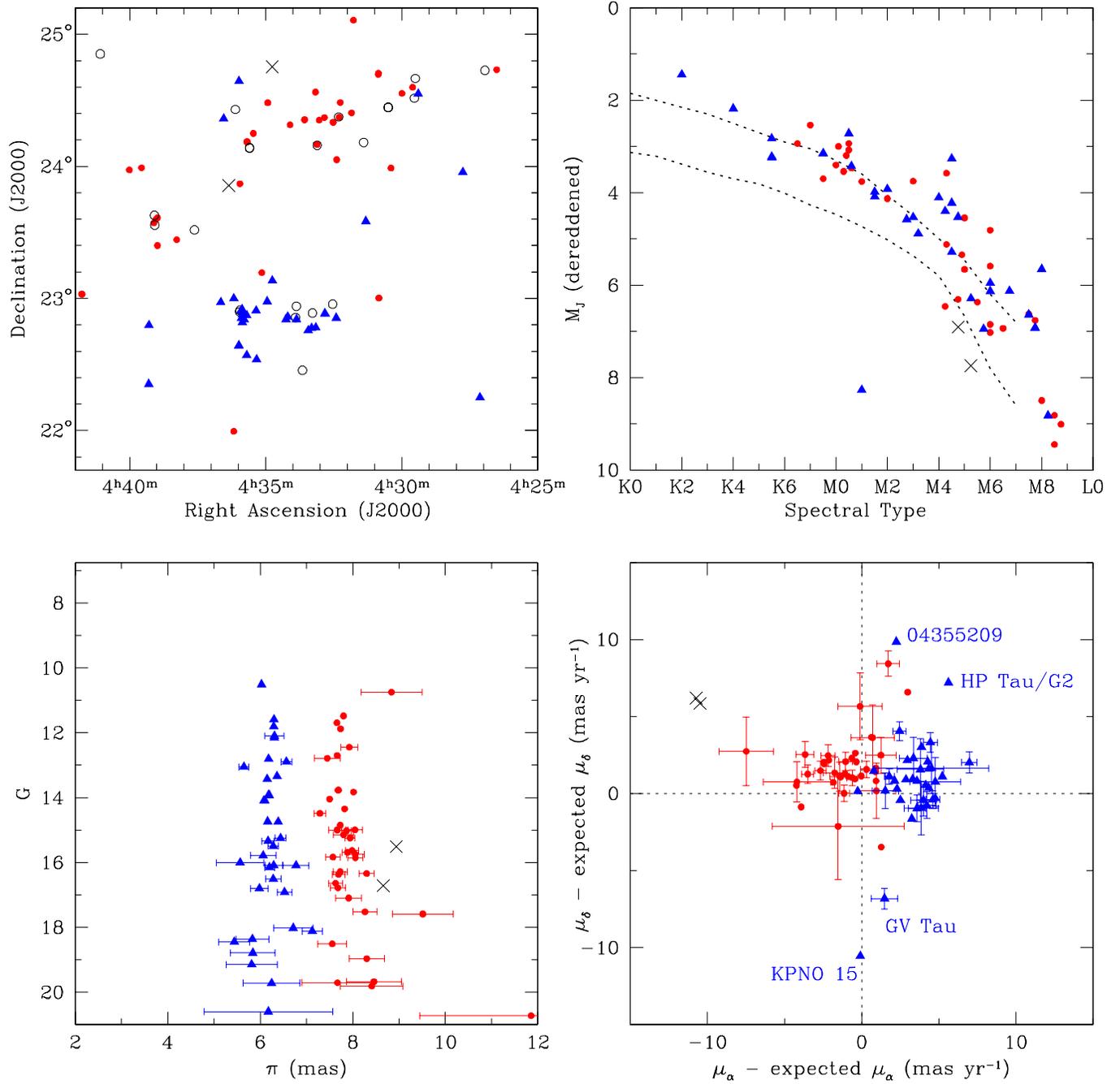}
\caption{
Same as Figure~\ref{fig:map1} for the L1524, L1529, and L1536 clouds.
}
\label{fig:map4}
\end{figure}

\begin{figure}
\epsscale{1.2}
\plotone{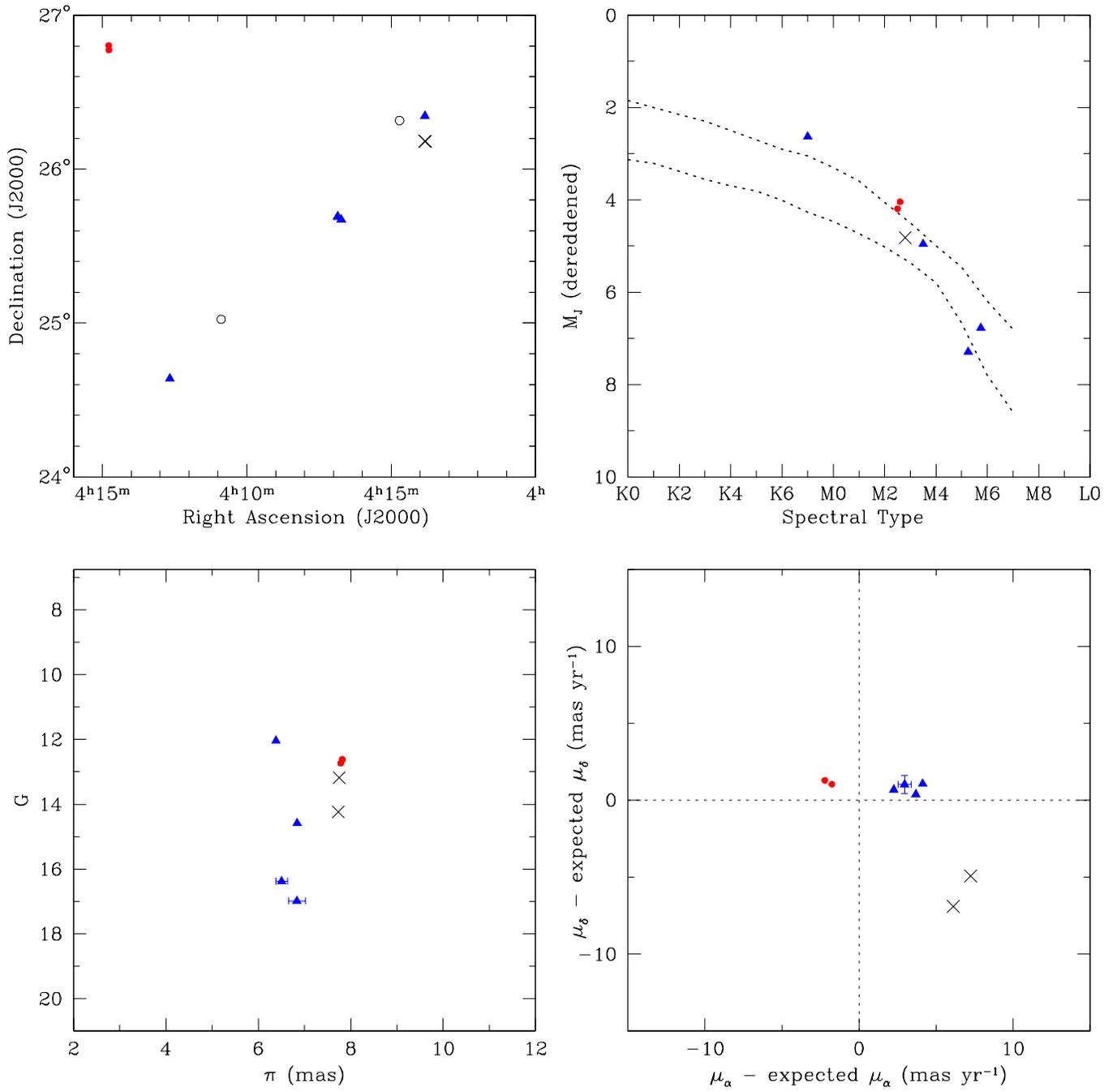}
\caption{
Same as Figure~\ref{fig:map1} for the L1489 and L1498 clouds.
}
\label{fig:map12}
\end{figure}

\begin{figure}
\epsscale{1.2}
\plotone{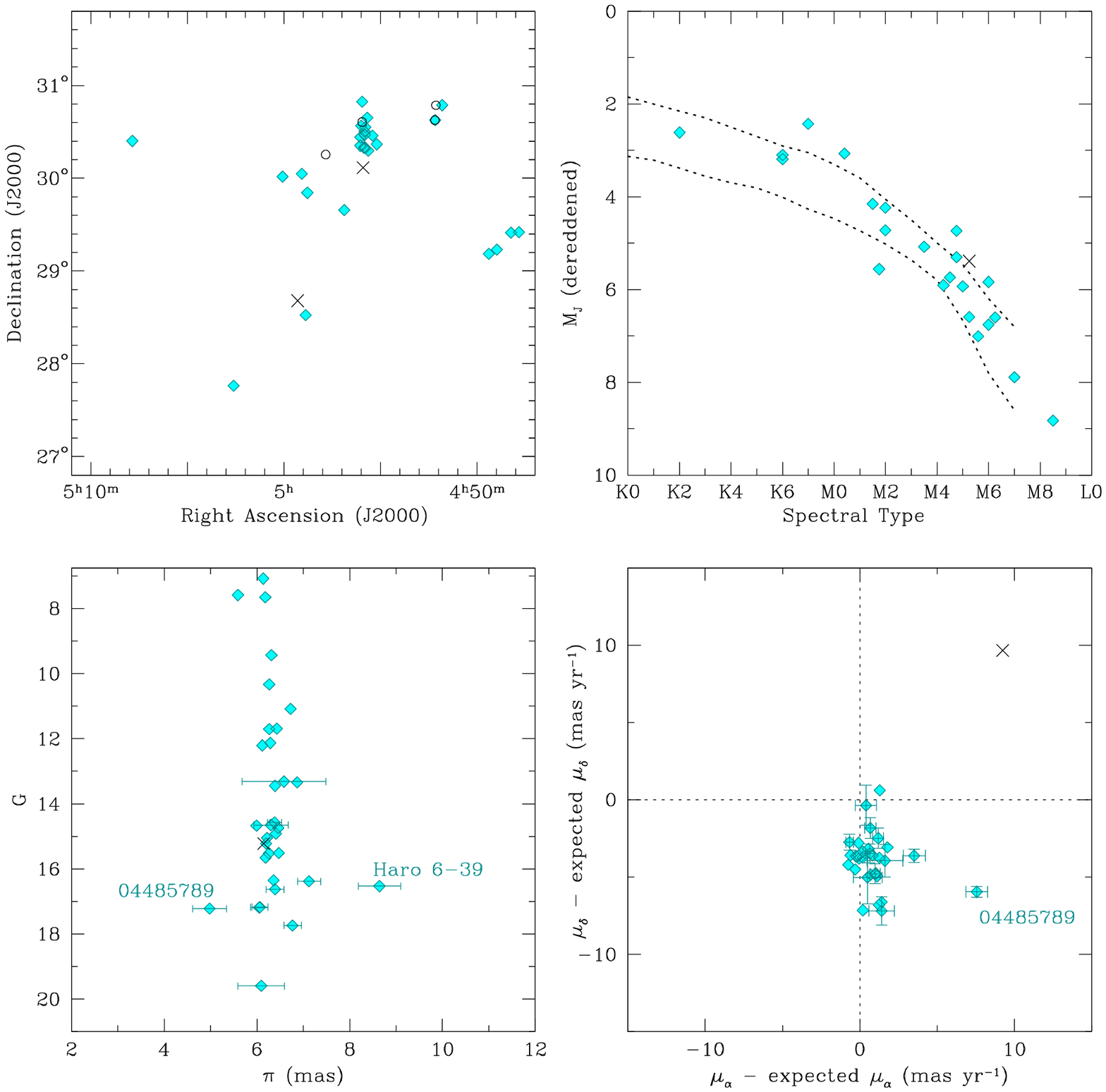}
\caption{
Same as Figure~\ref{fig:map1} for the L1517 cloud.
}
\label{fig:map10}
\end{figure}

\begin{figure}
\epsscale{1.2}
\plotone{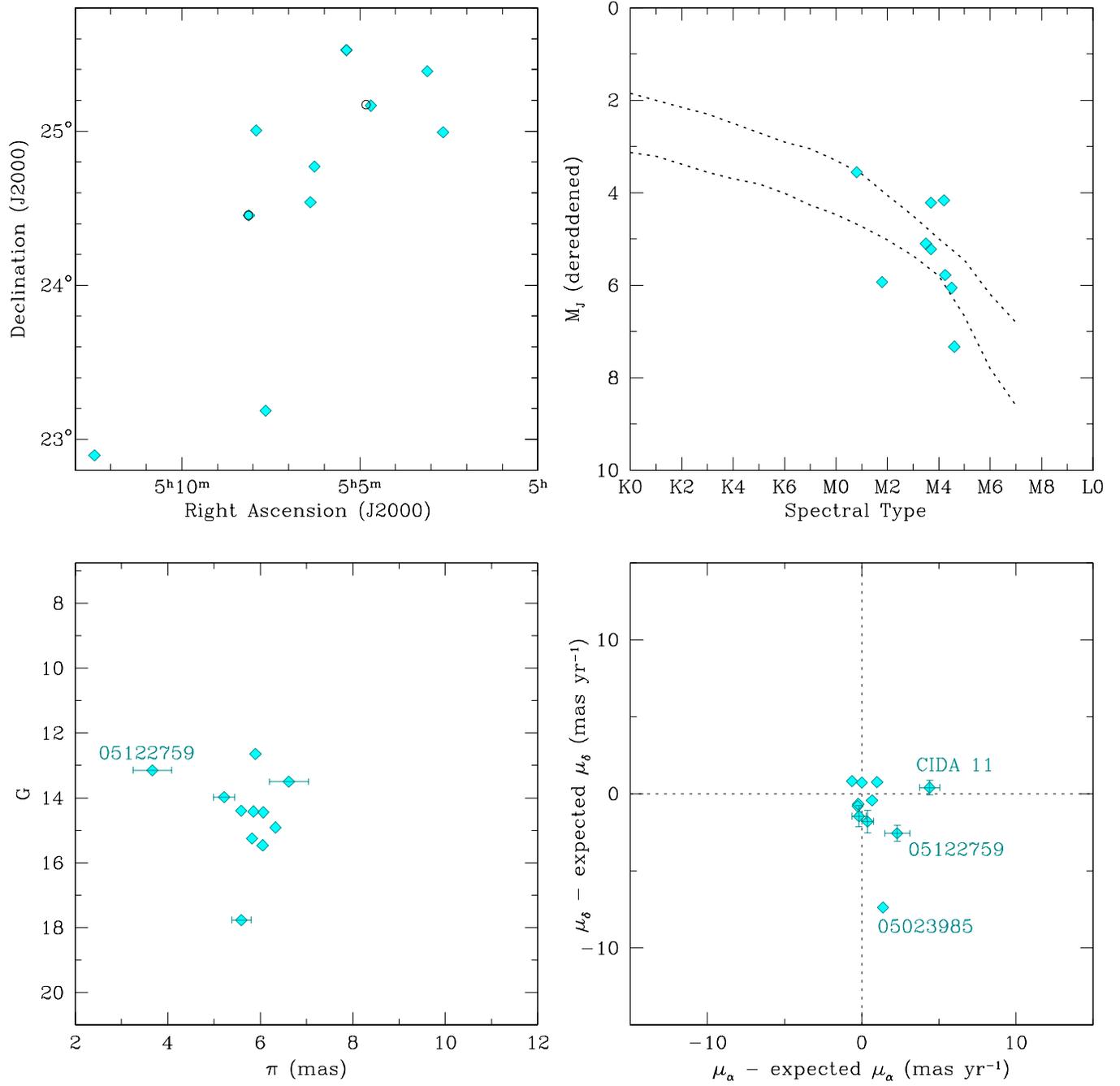}
\caption{
Same as Figure~\ref{fig:map1} for the L1544 cloud.
}
\label{fig:map11}
\end{figure}

\begin{figure}
\epsscale{1.2}
\plotone{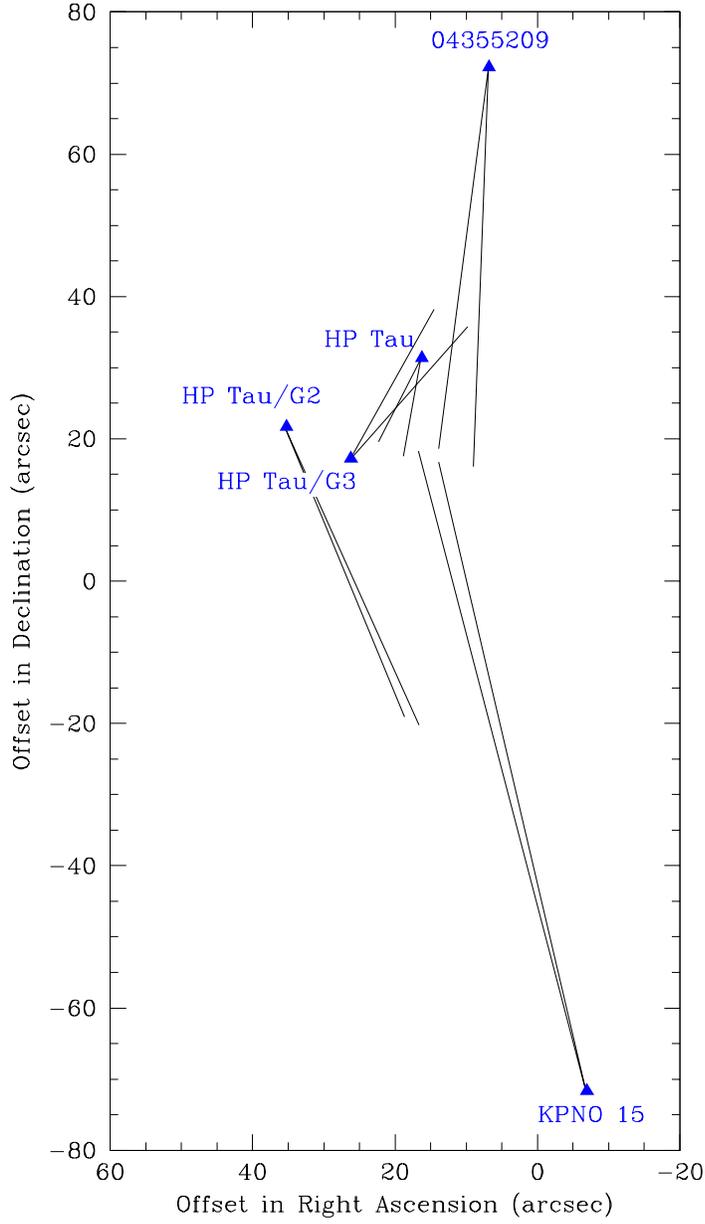}
\caption{
Relative positions of two stars in the L1536 cloud with discrepant proper
motions (Fig.~\ref{fig:map4}), 2MASS J04355209+2255039 and KPNO~15, and
other young stars in their vicinity. Based on proper motions from {\it Gaia}
DR2, those two stars were near the same location $\sim7200$ years ago.
The allowed paths over that time period are indicated for all
stars ($1\sigma$, lines), which are computed using the motions
relative to the average motion of the stars in this field.}
\label{fig:ejected}
\end{figure}

In Table~\ref{tab:mem}, I have included the proper motions and parallaxes from 
{\it Gaia} DR2, radial velocities from previous studies, $UVW$ space velocities 
computed from the {\it Gaia} data and radial velocities \citep{joh87}, the 
three bands of {\it Gaia} photometry, and the color codes for the kinematic
populations described in the previous section. 
A few of the radial velocity measurements lack estimates of errors. 
In those cases, an error of 1~km~s$^{-1}$ has been adopted when calculating 
the $UVW$ errors.  I have not used the radial velocity measurement for
HN~Tau~A from \citet{ngu12} because its systematic noise is large.
Systematic errors in the {\it Gaia} DR2 parallaxes are expected to be
less than 0.1~mas \citep{bro18}. Recent studies have found that those
parallaxes may be too small by $\sim0.08$~mas on average \citep{kou18,sta18}.
Since such errors may vary with position on the sky and their average value in 
the direction of Taurus is unknown, no correction has been applied to the 
parallaxes when deriving $UVW$ velocities. The latter have not been computed 
for stars that have discrepant parallaxes based on the analysis in the 
previous section. Table~\ref{tab:mem} contains estimates of $UVW$ for 100 
stars. The median of those velocities is 
$U, V, W = -15.9, -12.4, -9.4$~km~s$^{-1}$, which is similar to values from
\citet{ber06} and \citet{luh09tau}. That median $UVW$ was used in the
calculation of the proper motion offsets in 
Figures~\ref{fig:map1}--\ref{fig:map15}.

In Table~\ref{tab:pmpi}, I have compiled the medians of the parallaxes, proper 
motions, and proper motion offsets and the standard deviations of the
proper motions for each of the fields and populations in
Figures~\ref{fig:map1}--\ref{fig:map11}. Only stars with parallax errors of
$\leq$10\% and non-discrepant parallaxes have been considered.
If standard deviations are calculated for proper motions
with errors of $<0.25$~mas~yr$^{-1}$,
the most compact aggregates (those in B209, L1495, L1529, L1536, L1551, L1517)
have one-dimensional dispersions of $\sim1$~mas~yr$^{-1}$, which
corresponds to $\sim0.7$~km~s$^{-1}$ at the distances of the stars.
Since these values are significantly larger than the proper motion errors,
they should be dominated by the kinematics within the aggregates.

\begin{figure}
\epsscale{1.2}
\plotone{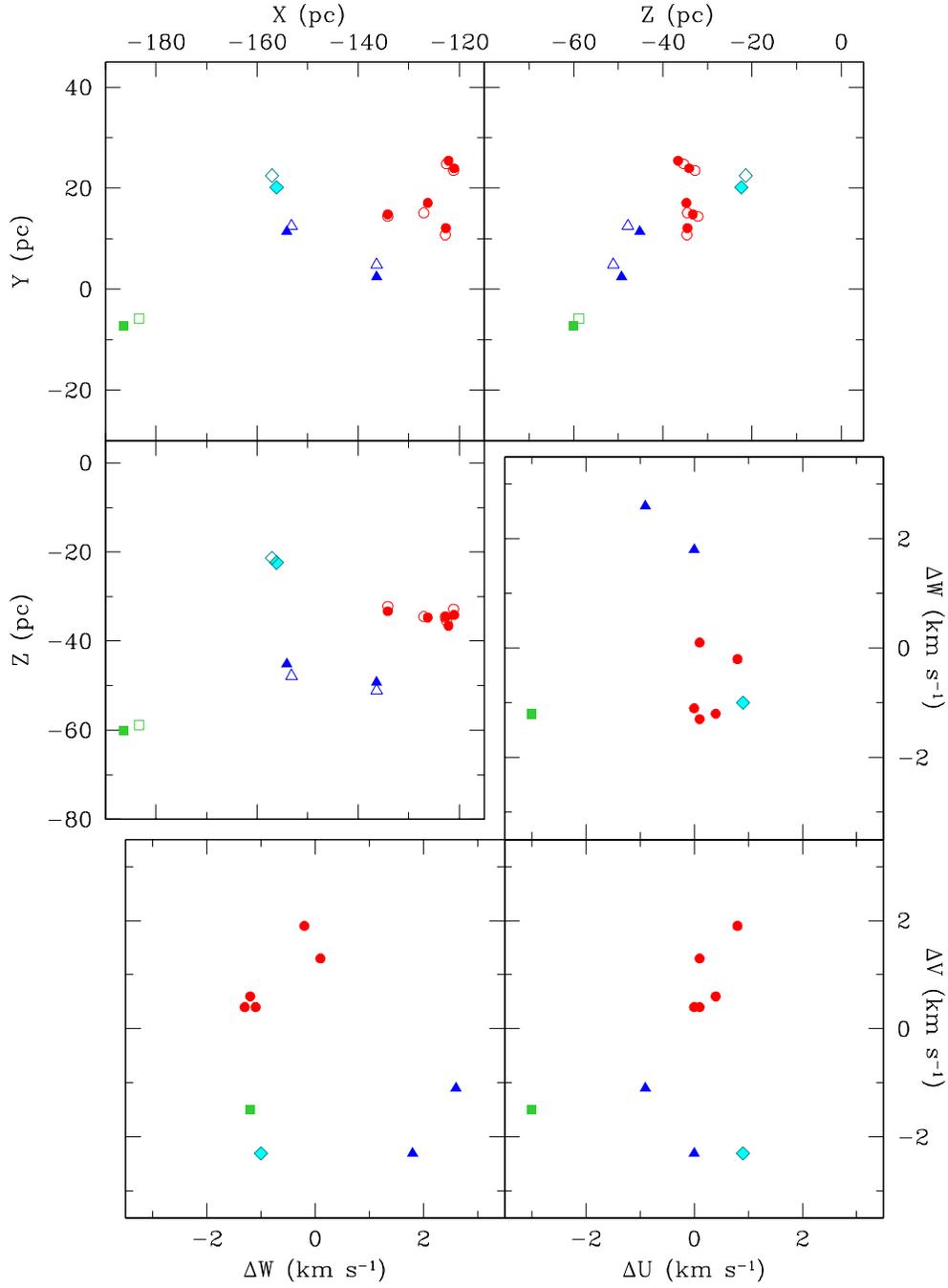}
\caption{
Positions and velocity offsets in Galactic Cartesian coordinates for 
aggregates of stars in Taurus (filled symbols, Table~\ref{tab:uvw}).
In the diagrams of $XYZ$, the open symbols represent the projected
positions at 1~Myr in the past. The velocity offsets are relative to
the median $UVW$ of known Taurus members.}
\label{fig:xyz}
\end{figure}

For groups of 3 or more stars that are near clouds and that have measurements
of $UVW$, I have calculated the medians of the parallaxes, proper motions, and
$UVW$ and the standard deviations of $UVW$.
The results are listed in Table~\ref{tab:uvw} with their associated clouds.
The dispersions are larger in $U$ than $V$ and $W$ because 
the radial velocities generally have larger errors than the proper motions.
In Figure~\ref{fig:xyz}, I have plotted the corresponding $XYZ$ positions
in Galactic Cartesian coordinates and the median $UVW$'s relative
to the median value for Taurus. 
The five aggregates labeled as red have similar velocities,
which is not surprising given their proper motion offsets
(Figs.~\ref{fig:map1}--\ref{fig:map4}).
The two aggregates labeled as blue, L1536 and L1551, have similar motions
and differ from the red aggregates by $\sim3$~km~s$^{-1}$ in each of
the $V$ and $W$ components.
The small aggregate associated with L1558 (green) differs by 
$\sim3$~km~s$^{-1}$ from the other groups in $U$ and is similar to the blue 
and red groups in $V$ and $W$, respectively.  It is also the most distant
aggregate in Taurus at nearly 200~pc, as mentioned in Section~\ref{sec:map13}. 
The cyan group near L1517 is similar to the blue/green and red groups in 
$V$ and $W$, respectively.  The total spread among the aggregates is
$\sim3$~km~s$^{-1}$ in each of the velocity components.
Those relative motions correspond to $\sim3$~pc ($\sim1\arcdeg$) in 1~Myr,
or $\sim10$\% of the diameter of the cloud complex, as illustrated in
Figure~\ref{fig:xyz}.

\section{Previous Candidate Members at Older Ages}
\label{sec:oldsamples}

Several studies over the last 30 years have proposed the existence of stars
with ages of $\gtrsim10$~Myr that are associated with the Taurus clouds
(Section~\ref{sec:intro}). In this section, I use
astrometry from {\it Gaia} DR2 to assess the membership of such stars
from \citet{kra17} and \citet{zha18}.

\subsection{Candidates from \citet{kra17}}
\label{sec:kra17}

\citet{kra17} compiled a catalog of 396 diskless stars that had been previously
identified as possible members of Taurus. Through analysis of several
diagnostics (e.g., proper motions, radial velocities, spectroscopic
signatures of youth), they concluded that 218 of the candidates were confirmed
or likely members. Roughly 1/3 of those suggested members were absent
from earlier compilations, most of which were older
($\gtrsim10$~Myr) and more widely distributed than the canonical members.
\citet{kra17} proposed that these stars represent an earlier generation of star
formation associated with the Taurus cloud complex.

Among the 218 stars that \citet{kra17} designated as members,
82 were absent from the census in \citet{luh17}.
\citet{esp17} examined the astrometric evidence of membership for those 82
stars. Sixteen of them had measurements of parallaxes and proper
motions from {\it Gaia} DR1 and appeared between
$\alpha=4^{\rm h}$--$5^{\rm h}10^{\rm m}$ and $\delta=15$--$31\arcdeg$,
which corresponds roughly to the boundaries of Figure~\ref{fig:map} and
encompasses all of the Taurus clouds.
\citet{esp17} compared the proper motions, parallaxes, and $M_J$ for
those stars and members from \citet{luh17} that had {\it Gaia} DR1 data.
Most of the former were kinematically distinct from the latter and exhibited
older ages ($\gtrsim$10~Myr).
The two samples differed by $\sim$10~mas~yr$^{-1}$ on average,
which corresponds to a relative drift of nearly $30\arcdeg$ over 10~Myr,
indicating that they are physically unrelated.

\begin{figure}
\epsscale{1.2}
\plotone{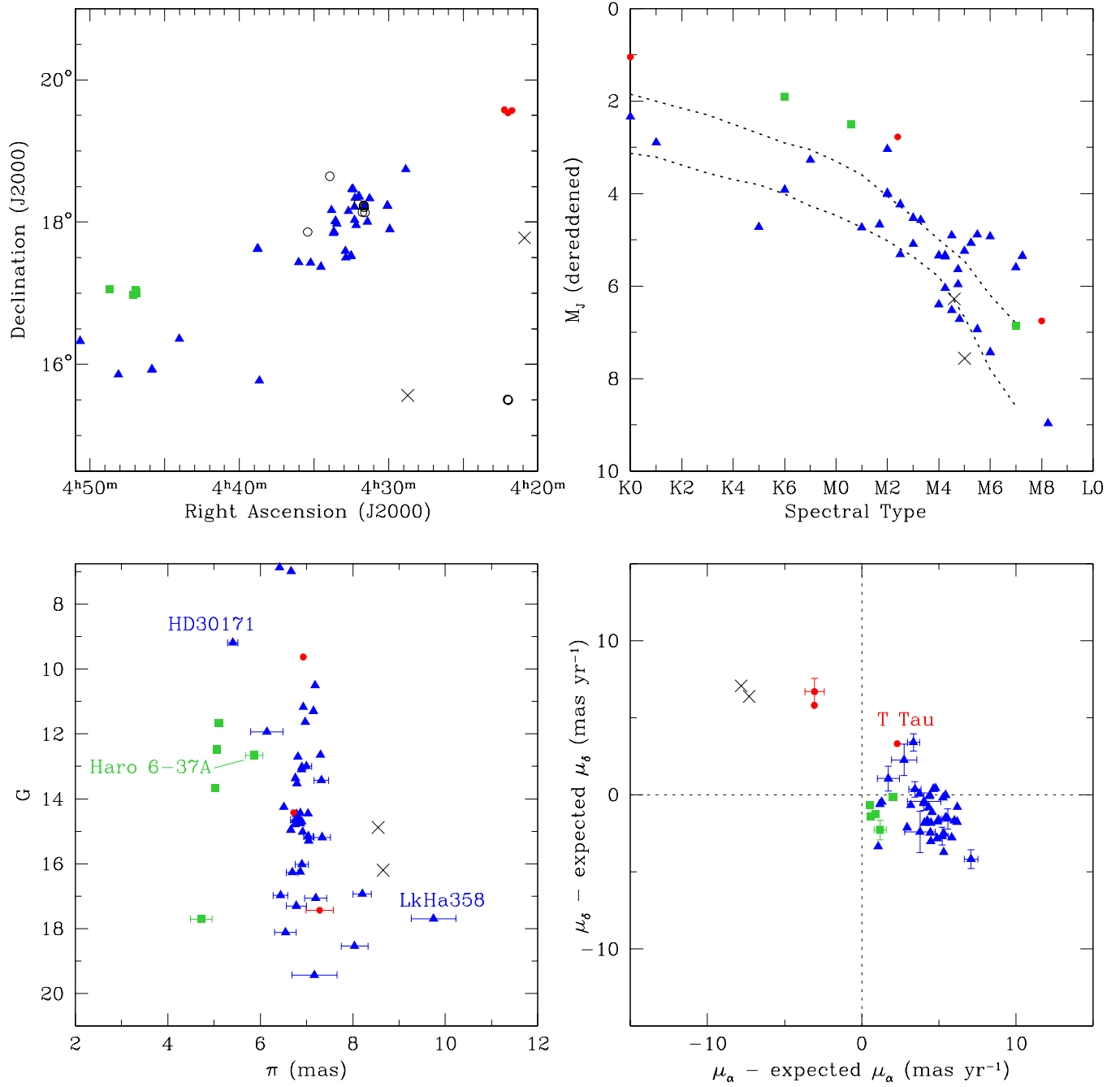}
\caption{
Same as Figure~\ref{fig:map1} for the L1551 and L1558 clouds and the small
cloud near T Tau.
}
\label{fig:map13}
\end{figure}

\begin{figure}
\epsscale{1.2}
\plotone{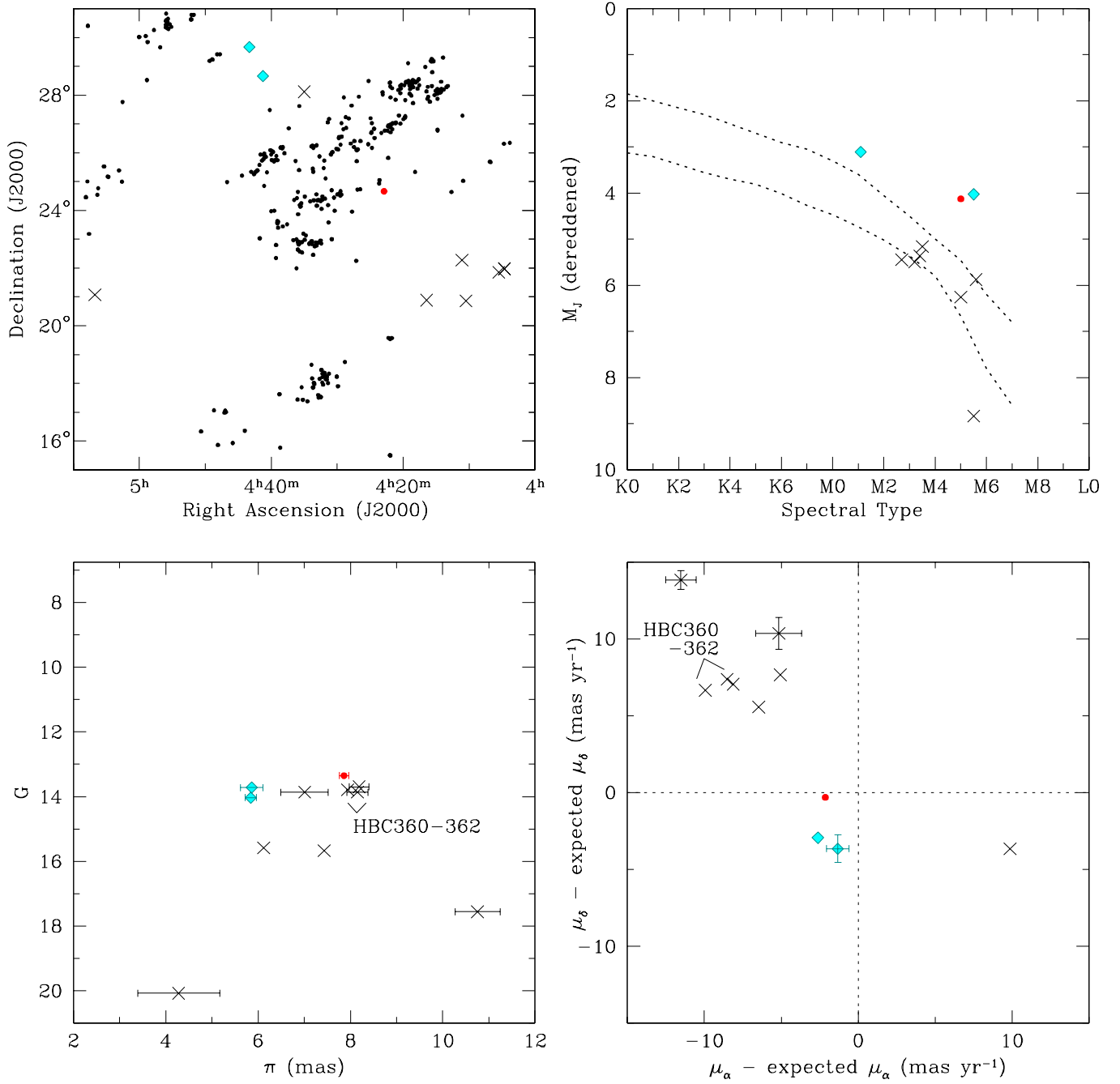}
\caption{
Same as Figure~\ref{fig:map1} for the area outside of the fields in
Figures~\ref{fig:map1}--\ref{fig:map11}.
}
\label{fig:map15}
\end{figure}

\begin{figure}
\epsscale{1.2}
\plotone{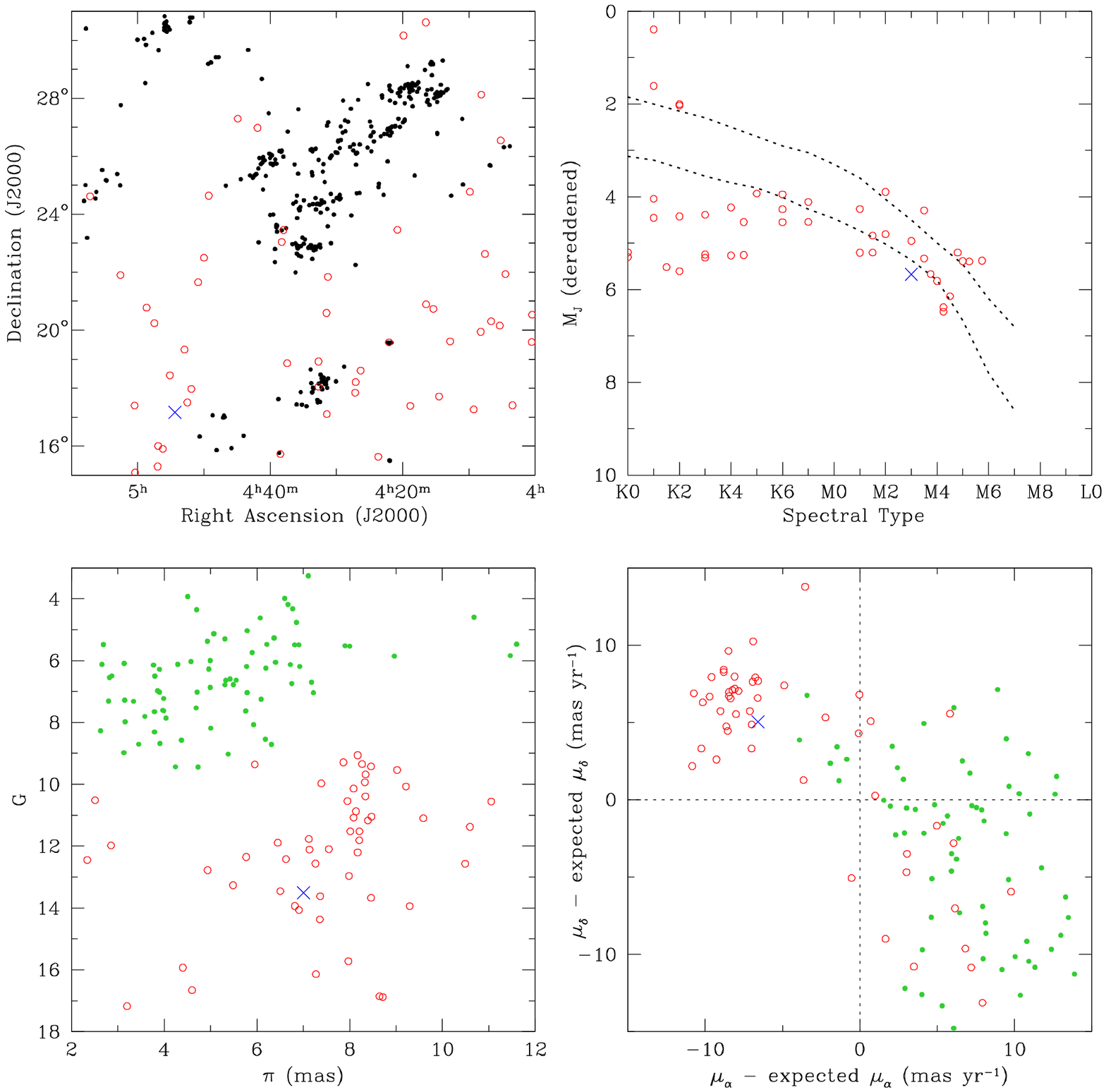}
\caption{
Same as Figure~\ref{fig:map1} for candidate members of Taurus from 
\citet{kra17} that were not adopted as members by \citet{esp17} (red open 
circles).
The map includes the stars compiled by \citet{esp17} that are adopted
as Taurus members in this work (filled circles; Section~\ref{sec:revised})
and the two bottom diagrams include proposed members of Cas-Tau from
\citet{dez99} (green filled circles). The young star St34 is also shown in each
of the diagrams \citep[blue cross,][]{har05st,whi05}.
}
\label{fig:map14}
\end{figure}

{\it Gaia} DR2 enables a comprehensive analysis of the candidate members
from \citet{kra17}.
The compilation of members from \citet{esp17} does not contain 
85 of the 218 stars identified by \citet{kra17} as probable members.
Fifty-two of those 85 stars are within the field defined by
$\alpha=4^{\rm h}$--$5^{\rm h}10^{\rm m}$ and $\delta=15$--$31\arcdeg$
and have parallaxes from {\it Gaia} DR2 with errors of $\leq10$\%.
They are plotted in diagrams in Figure~\ref{fig:map14} like those
in Figures~\ref{fig:map1}--\ref{fig:map15}.
Four stars have parallaxes and proper motions offsets that overlap with
the populations of members in Figures~\ref{fig:map1}--\ref{fig:map15},
consisting of L1551-55, RX~J0507.2+2437, RX~J0422.1+1934, and HD~283782.
The first two stars were in the census from \citet{luh17} but were
rejected by \citet{esp17}. I adopt these four stars as members.
One star, HBC~392, is somewhat close to the distribution of parallaxes
and proper motion offsets for one of the Taurus groups, L1551.
It has unusually weak Li absorption for a Taurus member \citep{wal88},
which has been cited as evidence that it is a nonmember \citep{har03}
or a Li-depleted member \citep{ses08}. HBC~392 appears below the median
sequence for Upper Sco \citep[11~Myr,][]{pec12} in $M_J$ versus spectral type,
indicating that its weak Li is a reflection of an older age.
Based on its {\it Gaia} astrometry and its radial velocity
\citep{ngu12}, it has a space velocity of
$U, V, W = -14.4\pm0.1, -16.1\pm0.1, -9.2\pm0.1$~km~s$^{-1}$, which differs 
from the median velocity of L1551 (Table~\ref{tab:uvw}) by
$\sim1.5$~km~s$^{-1}$ in each component. 
That difference corresponds to a relative drift of $>$20~pc ($>9\arcdeg$)
on the plane of the sky since the star was born, assuming an age of $>10$~Myr.
Thus, it is unlikely that HBC~392 originated in that cloud, and it is not
adopted as a Taurus member in this work.

\citet{har91} suggested that a previous sample of $\gtrsim10$~Myr stars 
towards Taurus from \citet{wal88} belong to the Cas-Tau association,
whose proposed members encompass the Taurus clouds and extend well beyond
them \citep{bla56}. Therefore, I have considered that possibility for the
47 stars in Figure~\ref{fig:map14} that differ kinematically from
the Taurus populations.
\citet{dez99} identified a sample of 83 B and A stars that may be members of
Cas-Tau. {\it Gaia} DR2 has provided parallaxes with errors of $\leq10$\% 
for 80 stars in that sample. Radial velocity measurements with errors of
$<4$~km~s$^{-1}$ are available for 33 of those 80 stars \citep{deb12b}.
The radial velocities combined with the {\it Gaia} astrometry produce
space velocities that have a median value of 
$U, V, W = -15.3, -22.0, -7.3$~km~s$^{-1}$, which is similar to the values
derived prior to {\it Gaia} DR2 \citep{dez99,dav18}.
The median velocity of Cas-Tau differs significantly from that of Taurus
($U, V, W = -15.9, -12.4, -9.4$~km~s$^{-1}$), as noted by \citet{dez99}.
The 80 proposed Cas-Tau members from \citet{dez99} that have {\it Gaia} data
are included in the bottom diagrams in Figure~\ref{fig:map14}.
Their parallaxes and proper motion offsets overlap with those of roughly
half of the 47 stars from \citet{kra17} that are kinematically distinct
from Taurus. Thus, it is plausible that the latter are members of Cas-Tau.

\begin{figure}
\epsscale{1.2}
\plotone{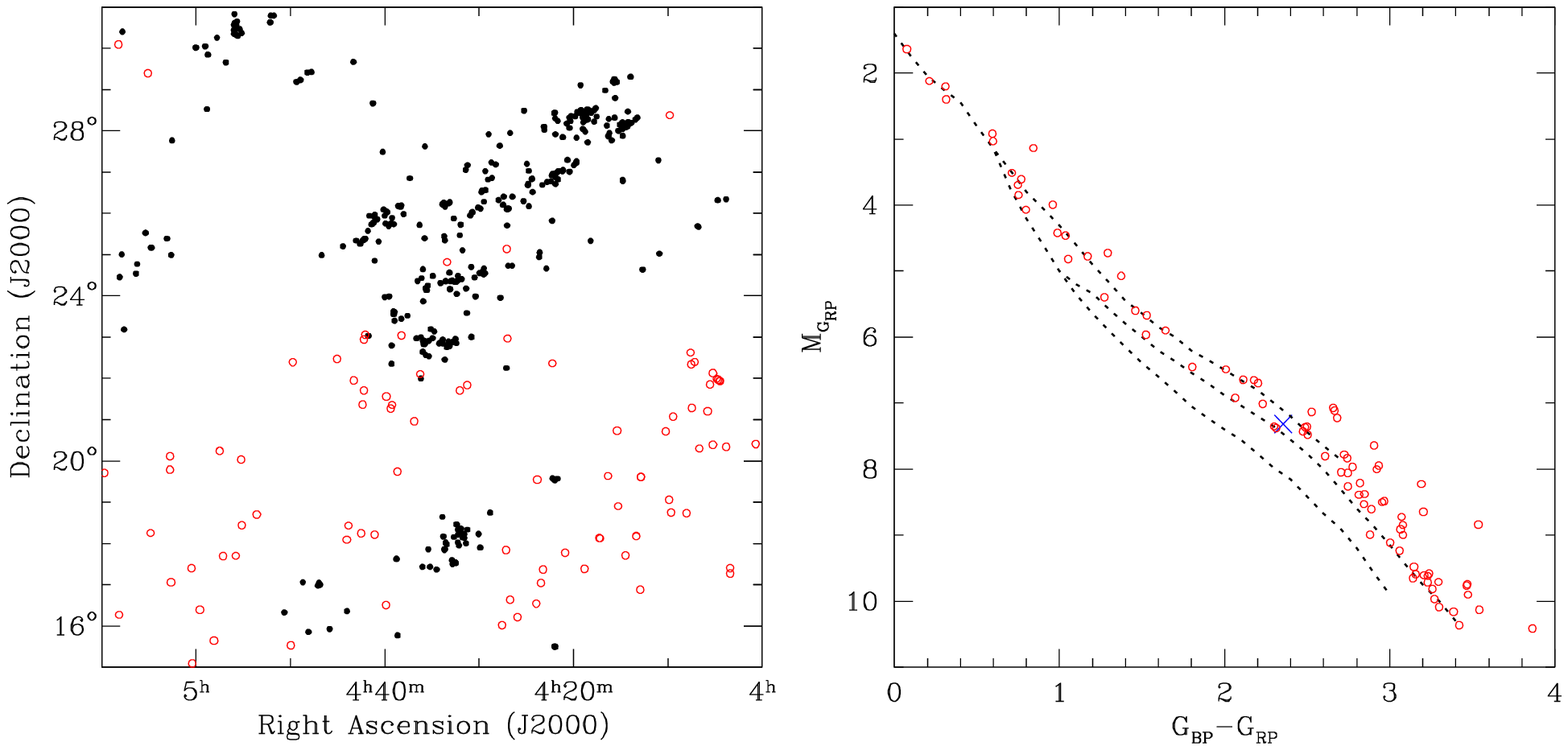}
\caption{
Spatial distribution and $M_{G_{\rm RP}}$ versus $G_{\rm BP}-G_{\rm RP}$ for
stars towards Taurus that have parallaxes and proper motion offsets
similar to those of the clump of candidates from \citet{kra17} at
$\pi\sim8.25$~mas and
($\Delta\mu_\alpha,\Delta\mu_\delta\sim-8.2,7.0$~mas~yr$^{-1}$) in
Figure~\ref{fig:map14} (red open circles).
The map includes the stars compiled by \citet{esp17} that are adopted
as Taurus members in this work (filled circles; Section~\ref{sec:revised})
and the color-magnitude diagram includes fits to the single-star sequences
for the $\beta$~Pic moving group \citep[24~Myr,][]{bel15},
the Tuc-Hor association \citep[45~Myr,][]{bel15}, and the Pleiades cluster
\citep[112~Myr,][]{dah15} (dotted lines, top to bottom). The young star St34
is also shown in the right diagram \citep[blue cross,][]{har05st,whi05}.
}
\label{fig:main}
\end{figure}

Most of the remaining stars in Figure~\ref{fig:map14} that do not overlap
with Cas-Tau form a clump in parallax and proper motion offset that is
centered near 8.25~mas and $(-8.2,7.0$~mas~yr$^{-1}$), respectively.
This clump coincides with group 29 from \citet{oh17}, which is
a possible new association of nine stars found with {\it Gaia} DR1.
Several of the stars from the catalog of members in \citet{esp17}
that were rejected in Section~\ref{sec:pops} also appear in that clump
(see Figs.~\ref{fig:map4}, \ref{fig:map13}, \ref{fig:map15}).  
To investigate the nature of this group, I selected stars 
from {\it Gaia} DR2 that are within 0.35~mas and 2~mas~yr$^{-1}$ of the
clump's center, which are the values beyond which the number of 
stars rapidly decreases. All of the nine stars identified by \citet{oh17} as
members of group 29 satisfy that proper motion threshold, but four of them
fall outside of the parallax threshold.
To allow for the possibility that members of the group might extend
beyond the confines of Taurus, I considered the area between 
$\alpha=3^{\rm h}$--$6^{\rm h}$ and $\delta=4$--$40\arcdeg$.
Most of the stars in the resulting sample (91/107) are within
the boundary of the map of Taurus in Figure~\ref{fig:map14}, so the
following discussion is restricted to those stars.
Their spatial distribution is shown in the map in Figure~\ref{fig:main}.
Measurements of radial velocities are available for 23 of the 91 stars
\citep[][{\it Gaia} DR2]{wal88,ngu12,kra17}. The velocities have a dispersion
of $\sim1.4$~km~s$^{-1}$, which is only somewhat larger than than the
dispersion of velocities on the plane of the sky imposed by the selection
criteria ($\sim0.5$~km~s$^{-1}$).
The median space velocity for those 23 stars is
$U, V, W = -13.0, -6.4, -9.7$~km~s$^{-1}$, differing by a total
of 6.6~km~s$^{-1}$ from the median motion of Taurus.

The ages of the 91 candidate members of group 29
can be estimated with a diagram of absolute magnitude
versus color. Since {\it Gaia} photometry has very high
precision, the diagram has been constructed with $G_{\rm BP}$ and $G_{\rm RP}$,
as shown in Figure~\ref{fig:main}.
A comparison of those stars to members of the Pleiades \citep{sta07}
in $G_{\rm RP}-K_s$ versus $J-H$ indicates that they have little extinction,
so extinction corrections have not been applied to the photometry.
In Figure~\ref{fig:main}, the stars form a sequence that is fairly narrow and
well-defined, which suggests that they comprise a coeval population.
To estimate the age of this sample, I have compared its sequence to those
of nearby clusters and associations that span a range of ages
\citep[][references therein]{bel15,gag18}. In Figure~\ref{fig:main},
I have included fits to the single-star sequences for three 
populations that bracket the sample, consisting
of the $\beta$~Pic moving group \citep[24~Myr,][]{bel15},
the Tuc-Hor association \citep[45~Myr,][]{bel15}, and the Pleiades cluster
\citep[112~Myr,][]{dah15}. The fits are defined in Table~\ref{tab:fits}.
This comparison suggests that group 29 is
slightly younger than Tuc-Hor ($\sim40$~Myr).
The velocity offset of 6.6~km~s$^{-1}$ relative to the Taurus combined
with an age of 40~Myr corresponds to a relative drift of 260~pc,
indicating that the stars have no relationship to the gas that would
eventually form the Taurus clouds.
Like Cas-Tau and the Hyades, group 29 is another example of a 
stellar population that lies in the direction of Taurus but is
unrelated to the cloud complex.
Astrometry and photometry for the 91 candidate members of group 29
are presented in Table~\ref{tab:cand2}.
Among these stars, HD~284149 and HBC~376 (TAP~26) are known to harbor
a brown dwarf companion and a hot Jupiter \citep{bon14,yu17}, respectively.

It was not included in the sample of stars analyzed by \citet{kra17}, but
St34 was cited in that study as an example of an old member of Taurus
($\gtrsim20$~Myr). 
\citet{whi05} found that it is a spectroscopic binary in which the components
have similar luminosities and spectral types.
It appeared to reside in Taurus based on its kinematics and its
evidence of youth in the form of an accretion disk, 
but the components lacked Li absorption, indicating an age of $>20$~Myr.
\citet{whi05} concluded that St34 is probably a relatively old
member of Taurus ($\gtrsim8$~Myr). Meanwhile, \citet{har05st}
proposed that the system is not associated with the Taurus clouds,
and that instead it lies in the foreground at a distance of $\sim100$~pc,
which appeared to alleviate the discrepancy between the ages inferred
from the luminosity and the absence of Li absorption.
According to the parallax measurement
from {\it Gaia} DR2, St34 has a distance of 142.7$\pm1.2$~pc, which places
it within the range of distances of Taurus members.
However, the kinematics of St34 are inconsistent with membership in Taurus.
Based on the astrometry from {\it Gaia} and the radial velocity from 
\citet{whi05}, the system has a space velocity of 
$U, V, W = -15.1\pm0.7, -6.9\pm0.1, -10.7\pm0.1$~km~s$^{-1}$, which differs
by $\gtrsim4$~km~s$^{-1}$ from the median motions of the groups in Taurus
(Table~\ref{tab:uvw}). 
The data for St34 are included in both Figures~\ref{fig:map14}
and \ref{fig:main}. The photometry has been corrected for the binarity
by assuming that the components have equal fluxes.
For a single component, the spectral type and $M_J$ relative to the median
sequence of Upper Sco suggests an age of $\sim20$~Myr \citep{bar15}
while the position in the color-magnitude diagram 
relative to the $\beta$~Pic and Tuc-Hor associations indicates an age
of $\sim30$~Myr. The latter value could be overestimated if the system
has excess emission in $G_{\rm BP}$ from accretion. An age of 20--30~Myr
is consistent with the constraints on the Li abundance \citep{whi05}.

\subsection{Candidates from \citet{zha18}}
\label{sec:zha18}

\citet{zha18} presented a sample of 58 late-type objects that they
classified as members of Taurus. Most of them are fainter than the
known members at a given color or spectral type, indicating that they
are older or more distant. \citet{zha18} concluded that these objects
represent an older population ($\gtrsim10$~Myr) that is similar to the
one proposed by \citet{kra17}.

\begin{figure}
\epsscale{1.2}
\plotone{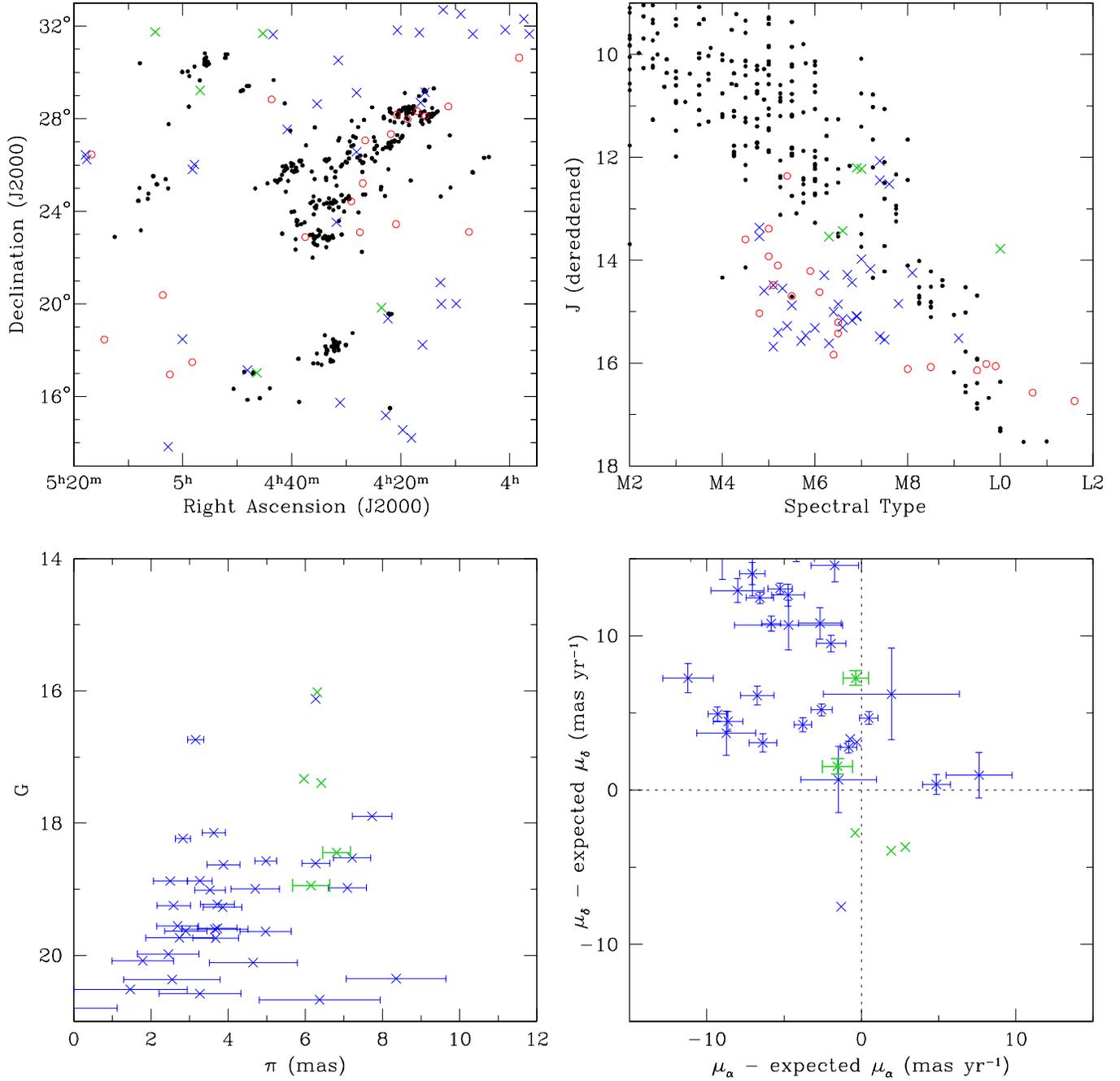}
\caption{
Same as Figure~\ref{fig:map1} for candidate members of Taurus from 
\citet{zha18} that have parallax measurements from {\it Gaia} DR2 (blue and
green crosses) and those that lack such data (red open circles). 
The five candidates adopted as members in this work are plotted in
green and the other candidates are plotted in blue.
The top diagrams include the stars compiled by \citet{esp17} that are adopted
as Taurus members in this work (filled circles; Section~\ref{sec:revised}).
}
\label{fig:zha}
\end{figure}

The membership of the candidates from \citet{zha18} can be assessed
with data from {\it Gaia} DR2, which became available after that study.
Among the 58 candidates, 47 have entries in {\it Gaia} DR2
and 38 have parallax measurements (15 with errors of $\leq10$\%).
In Figure~\ref{fig:zha}, all of the candidates are plotted on a map of Taurus
and a diagram of extinction-corrected J versus spectral type \citep{zha18}.
Those diagrams also include the stars from \citet{esp17} that are
adopted as members in this work.
The stars with parallax measurements are shown in diagrams of
$G$ versus parallax and proper motion offsets relative to the motion
expected for the median space velocity of known Taurus members.
For those offsets, I have adopted the parallactic distances when the parallax
errors are $\leq10$\% and otherwise have assumed a distance of 140~pc.

Five of the 38 candidates with {\it Gaia} parallaxes and proper motions
overlap with the groups of known members in those parameters, consisting
of PSO J065.8871+19.8386, PSO J071.3189+31.6888, PSO J071.6033+17.0281,
PSO J074.1999+29.2197, and PSO J076.2495+31.7503.
Based on those data and the evidence of youth in the spectra from \citet{zha18},
I have adopted them as members of Taurus.
Three of those five objects, PSO J065.8871+19.8386, PSO J071.6033+17.0281,
and PSO J074.1999+29.2197, have been independently identified as members
by \citet{esp18}.
Among the remaining 33 candidates with {\it Gaia} astrometry,
PSO J070.2057+27.5378 and PSO J079.3986+26.2455 are somewhat close to
the distributions of parallaxes and proper motion offsets for 
L1517 and L1544, respectively, but are located rather far from those clouds
($\sim3\arcdeg$).
Given that young stars unrelated to the Taurus clouds are scattered
across this area of sky (Section~\ref{sec:kra17}), those
two stars have insufficient evidence of membership.
The remaining 31 candidates with {\it Gaia} data have discrepant parallaxes
and proper motions (see Fig.~\ref{fig:zha}), and thus are excluded from my
catalog of members.
All of the five candidates from \citet{zha18} that have kinematics
consistent with membership appear within the Taurus sequence 
in the diagram of $J$ versus spectral type, indicating that they are within
the age range of the known members. None of the candidates for older
members with {\it Gaia} astrometry have been confirmed as such by those data.

{\it Gaia} parallaxes and proper motions are unavailable for 20 of the
candidates from \citet{zha18}.
Given the lack of astrometry with sufficient precision to
distinguish between Taurus members and young contaminants, I assess
those candidates with the ages implied by the color-magnitude diagram
in Figure~\ref{fig:zha} and the proximity to the Taurus clouds.
The use of age as a criterion is justified by a search
of {\it Gaia} DR2 for undiscovered members at higher masses in
Section~\ref{sec:search}, which demonstrates that a population older than
the known members does not exist.
Two of the 20 candidates that lack {\it Gaia} astrometry,
PSO J064.6887+27.9799 and PSO J065.1792+28.1767, are within the
sequence of known members in the diagram of $J$ versus spectral type
and are near the clouds. They were independently found and classified
as members by \citet{esp17}.
A few additional candidates like PSO J059.5714+30.6327 may be as young
as the known members, but they are far from the clouds and cannot be
reliably distinguished from young contaminants with the available data.
Most of the 20 candidates are too faint to be members that are coeval
with the known Taurus population, as shown in Figure~\ref{fig:zha}.

\begin{figure}
\epsscale{1.2}
\plotone{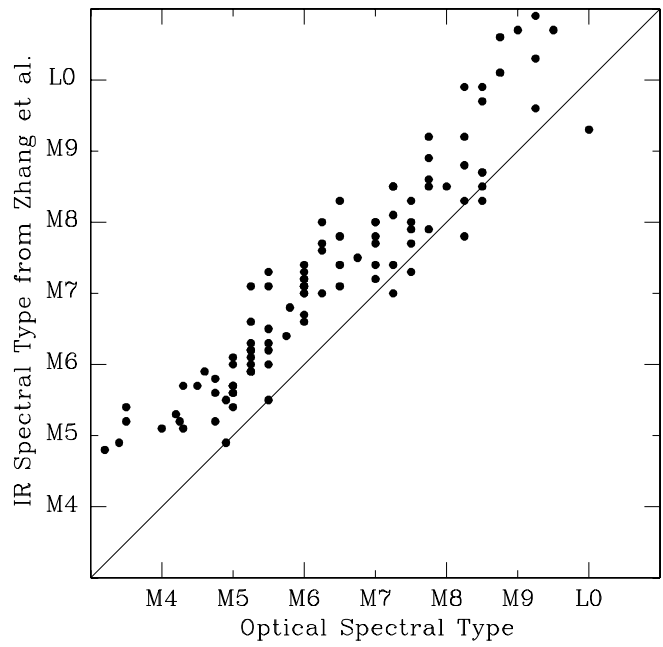}
\caption{
Near-IR spectral types from \citet{zha18} versus optical spectral types
from previous studies (see Section~\ref{sec:zha18}) 
for known members of Taurus that were classified in the former study.
}
\label{fig:spt}
\end{figure}

Among the seven candidates from \citet{zha18} that
are included in my catalog of members, five have been
spectroscopically classified by \citet{esp17,esp18},
who derived the following spectral types:
M9.25 (IR) for PSO J064.6887+27.9799 and J065.1792+28.1767,
M7 (optical) for PSO J071.6033+17.0281, M6 (optical) for PSO J074.1999+29.2197,
and M9 (optical/IR) for PSO J065.8871+19.8386.
I have measured a type of M5.5 for both of the remaining two stars,
PSO J071.3189+31.6888 and PSO J076.2495+31.7503, using the IR spectra
from \citet{zha18}.
For those seven stars, the classifications from \citet{zha18} are later than
those from \citet{esp17,esp18} and this work by an average of $\sim1$~subclass.
In addition to their candidates, \citet{zha18} classified
IR spectra of most known late-type members of Taurus.
In Figure~\ref{fig:spt}, those types are compared to the optical spectral types
that are available for those objects
\citep{bri98,bri02,mar99,hk03,whi03,gui06,sle06,luh04tau,luh06tau1,luh03tau,luh06tau2,luh09tau,esp14,her14}\footnote{
Most of these optical types were derived via comparison to
the average spectra of dwarf and giant standards \citep{hen94,kir91,kir97},
which is a scheme that has been applied to M5--M9.5 members of Taurus and
other star-forming regions during the past two decades
\citep[][references therein]{luh97,luh98vx,luh98,luh99,luh12}.}.
Once again, the classifications from \citet{zha18} are
systematically later by $\sim1$~subclass. Thus, their types cannot be used
alongside the previous optical types in a meaningful way.
The IR types from my previous studies are based on comparison to
optically-classified members of Taurus and other star-forming regions
\citep{luh17}, which are the ideal standards for producing IR types that
are on the same system as the optical types.

\section{Search for New Members}
\label{sec:search}

The data from {\it Gaia} DR2 can be used to search for
stars associated with the Taurus clouds with a high degree of
completeness for all locations, ages, and stellar masses 
with the exception of the most highly reddened members. Most of the latter
are likely to be younger and less evolved, and hence should have been found
by mid-IR surveys for stars with circumstellar disks 
\citep{bei86,ken90,luh06tau2,reb10,esp14}.

For each of the nine fields in Figure~\ref{fig:map}, I selected stars
that have measurements of parallaxes from {\it Gaia} DR2 with errors of
$\leq10$\% and that are within 0.5~mas and 4~mas~yr$^{-1}$ of the median
parallaxes and proper motion offsets of any of the populations of
known members within that field (Table~\ref{tab:pmpi}). 
These thresholds were selected to be large enough to recover most (95\%)
of the known members that have the necessary {\it Gaia} data and that do
not have discrepant parallaxes (Section~\ref{sec:pops}).
If larger thresholds are adopted, only a few additional candidates coeval
with the known Taurus population are selected while the number of candidates
older than Taurus increases roughly in proportion to the square of the
thresholds, which is consistent with a population of field contaminants.
Among the candidates selected by my criteria, 
I have assigned membership to those with previous spectroscopic data that
are consistent with membership and those that are within a few arcseconds
of known members, and hence are likely to be companions. They consist of
HD~28354, HD~30378, HD~283641, JH~223~B, V892~Tau~NE, XEST~20-071~B,
2MASS J04284263+2714039~B, and 2MASS J05080816+2427150~B.
I rejected candidates that have been previously classified as evolved stars
or that have radial velocities that differ significantly ($>5$~km~s$^{-1}$)
from the median velocities of the Taurus populations, which applies to
all (9) candidates for which velocities have been measured.
All of the candidates rejected by radial velocities are also much fainter
than the known members of Taurus at a given color, which further suggests
that they are field stars.
After these steps, there remain 141 candidate members, which have
magnitudes ranging from $G\sim13$--20.

\begin{figure}
\epsscale{1.2}
\plotone{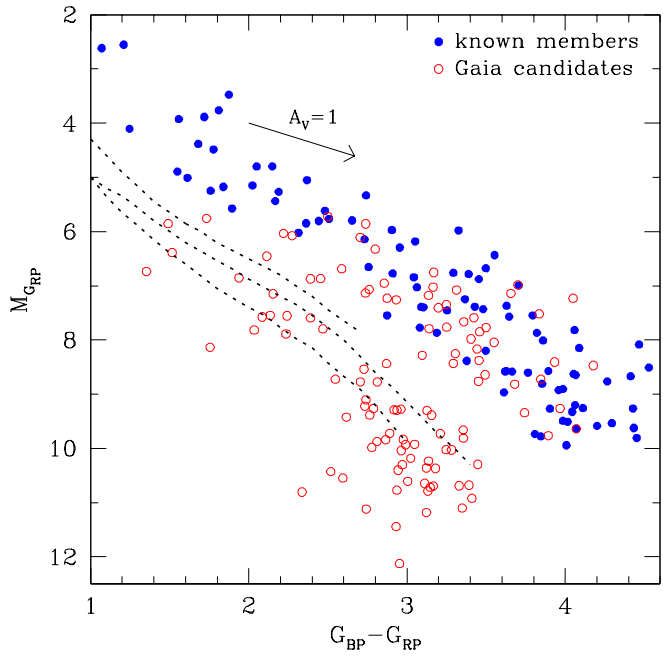}
\caption{
$M_{G_{\rm RP}}$ versus $G_{\rm BP}-G_{\rm RP}$ for known members of
Taurus that lack disks (blue filled circles) and candidate members 
in the nine fields in Figure~\ref{fig:map} that were selected from
{\it Gaia} DR2 to have parallaxes and proper motion offsets similar to 
those of the Taurus populations in those fields (red open circles).
I have included fits to the single-star sequences
for the $\beta$~Pic moving group \citep[24~Myr,][]{bel15},
the Tuc-Hor association \citep[45~Myr,][]{bel15}, and the Pleiades cluster
\citep[112~Myr,][]{dah15} (dotted lines, top to bottom).
}
\label{fig:br}
\end{figure}

To estimate the ages of the candidates, I have plotted the ones with
measurements of $G_{\rm BP}$ and $G_{\rm RP}$ (114 of the 141 candidates)
in a diagram of $M_{G_{\rm RP}}$ versus $G_{\rm BP}-G_{\rm RP}$ in
Figure~\ref{fig:br}. As done in Figure~\ref{fig:main}, I have included
fits to the single-star sequences for the $\beta$~Pic and
Tuc-Hor associations and the Pleiades cluster.
In an optical color-magnitude diagram, stars with disks occasionally appear
below the sequence for their population if their observed flux is dominated
by scattered light or if accreting material generates bright excess emission
at shorter wavelengths. Therefore, to more clearly define the sequence for
Taurus in Figure~\ref{fig:br}, only members that lack disks are shown
\citep{esp14,esp17}. Members with discrepant parallaxes are excluded
(Section~\ref{sec:pops}).

Many members of Taurus have substantial extinction, which affects their
locations in a color-magnitude diagram. 
Because the {\it Gaia} photometric bands are quite broad, the
relation between the extinction in a given band and the extinction at
a specific wavelength depends noticeably on the amount of extinction and
the intrinsic spectrum (or color) of the object \citep{bab18,dan18}.
A reddening vector that is applicable to typical members of Taurus 
is shown in Figure~\ref{fig:br}. Since the vector is largely parallel to
the Taurus sequence, the variable extinction among the members should not
broaden the sequence significantly.
Meanwhile, most of
the candidate members closely match the sequence of Pleiades members 
\citep{sta07} in color-color diagrams like $G_{\rm RP}-K_s$ versus $J-H$,
indicating that they have little extinction.
For these reasons, I have not attempted to correct the data in
Figure~\ref{fig:br} for extinction.

The candidates exhibit two distinct distributions in Figure~\ref{fig:br},
one that is scattered within the sequence of known Taurus members
and another that appears below the Tuc-Hor sequence ($\gtrsim40$~Myr).
None of the latter show evidence of disks in mid-IR photometry
from the {\it Spitzer Space Telescope} \citep{wer04} or the
{\it Wide-field Infrared Survey Explorer} \citep{wri10},
so their low positions in the diagram are not attributable to scattered light.
The sharp decrease in the number of members and candidates below the
lower envelope of the Taurus sequence indicates that that there are few,
if any, stars at ages of 10--40~Myr that are associated with the Taurus clouds.
Given the paucity of candidates in that age range, it is highly unlikely that
the stars at $>$40~Myr have any relationship to Taurus.
Indeed, most of the older candidates appear near the selection thresholds
for parallax and proper motion offsets or have larger astrometric errors,
whereas the younger candidates are more tightly clustered with the known
members in those parameters. 
In addition, the matching population for more than half of the older
candidates was L1558, which contains only five
known members. The unrealistically large number of candidates is likely
a reflection of the fact that this group is the most distant one in Taurus
($\sim200$~pc) and the number of stars satisfying the proper motion
criteria increases rapidly with larger distances.
To verify the plausibility that the older candidates comprise unrelated
contaminants, I performed multiple iterations of the selection of candidates
with uniform shifts applied to the median proper 
motion offsets of the Taurus groups (e.g., $\pm10$~mas~yr$^{-1}$). 
The resulting samples of stars closely resemble the older candidates selected 
for Taurus in size and distribution of colors and absolute magnitudes.

In Table~\ref{tab:cand}, I present the 54 candidates that have estimated
ages of $\lesssim20$~Myr. They consist of the 49 stars in 
Figure~\ref{fig:br} that appear above the sequence for $\beta$~Pic,
three candidates that have photometry in only one band ($G$) but that are
candidate companions to stars that are bright enough to appear in the
Taurus sequence, and two candidates that lack $G_{\rm BP}$ but are young
according to a diagram of $M_{G_{\rm RP}}$ versus $G-G_{\rm RP}$.
Spectroscopy of the candidates is necessary to measure their spectral types
and verify their youth. \citet{esp18} has classified spectra of many of
the candidates, all of which show evidence of youth that is consistent with 
the ages inferred from Figure~\ref{fig:br}. Most of those stars will
be adopted as members, but a few of them have motions that deviate enough
from those of the Taurus groups that they could be unrelated young
stars from Cas-Tau.

Each of the candidates from the preceding analysis was selected to reside
within one the nine Taurus fields and to have a similar parallax
and proper motion as one of the populations of known members within its field.
To search for candidates at larger distances from those populations, I have
identified stars at any location within Figure~\ref{fig:map}
that satisfy the previously applied thresholds of parallax and proper
motion for any of the populations of members.
These relaxed criteria produce an additional 51 candidates that appear
within the sequence of known members in color-magnitude diagrams.
Most of these candidates are far from the populations to which they were
matched ($>5\arcdeg$) and are near the thresholds for parallax and proper
motion, and thus are unlikely to be members.
The remaining (eight) candidates agree more closely with the astrometry for the
known populations and are within a few degrees of the boundaries of their
fields. The latter candidates have been included in Table~\ref{tab:cand}.

Four of the candidates comprise two $2\arcsec$ pairs, which correspond to
2MASS J04572852+3029107 and 2MASS J04355568+1707395. In addition,
2MASS J04161407+2758275, 2MASS J04291717+1826375, and 2MASS
J05010116+2501413 are $0\farcs9$, $1\farcs4$, and $1\farcs8$ pairs,
respectively, in which one component was selected as a candidate and the
other one was rejected by the criteria for parallax or proper motion. 
The rejected component in the first pair has a poor astrometric fit, perhaps
due to the binarity, and the rejected stars in the other two pairs
are only slightly beyond the thresholds for selection in proper motion.
2MASS J04411296+1813194 is a $2\farcs3$ pair in which one component is
a candidate and the other one lacks parallax and proper motion measurements
from {\it Gaia}.
Only the components of these various pairs that were identified as candidates
are listed in Table~\ref{tab:cand}.

\section{Properties of the Stellar Population}
\label{sec:stellarpop}

{\it Gaia} DR2 has made it possible to produce a highly refined census of
known members of Taurus and to perform a thorough search for
undiscovered members, which in turn should enable the best constraints
to date on the distributions of masses and ages in the region.

\subsection{Distribution of Ages}
\label{sec:ages}

Because low-mass stars ($\lesssim1$~$M_\odot$) are predicted to evolve
primarily in a vertical direction in the Hertzsprung-Russell diagram for at
least 10~Myr following their birth, the distribution of ages in a star-forming
region should be directly reflected in a spread in luminosities at a given
effective temperature. However, additional factors can contribute to the
observed spread in
luminosity estimates \citep{har01}, including unresolved binaries, uncorrected
emission from circumstellar material, variations in distances to the stars
(if a single distance is adopted for a population), uncertainties in
photometry, extinctions, and bolometric corrections, and differences 
in accretion histories \citep{bar09,lit11}.
Nevertheless, the luminosity spread in a star-forming region can provide
useful constraints on the distribution of ages \citep{bal99}.

For a given star, I use the offset in its extinction-corrected $M_J$ relative
to the median sequence for Taurus, $\Delta M_J=M_J-M_J$(median),
as a proxy for its relative age.
The $J$ band is selected for measuring the photospheric flux
as a compromise between shorter wavelengths where disk emission is lower
and longer wavelengths where extinction is lower.
I have computed $\Delta M_J$ for known members of Taurus
(Section~\ref{sec:revised}) that have spectral types between K0--M7,
estimates of extinction \citep{esp17,luh17}, and parallax measurements
from {\it Gaia} DR2 that have errors of $\leq10$\% and that are not discrepant
(Section~\ref{sec:pops}).
Stars with known edge-on disks, most protostars, and some close companions
lack extinction estimates or spectral classifications, and hence are excluded.
In addition, I have estimated $\Delta M_J$ for the candidate members 
identified in the previous section (Table~\ref{tab:cand}).
Each candidate was dereddened to the Pleiades locus in $J-H$ versus
$G_{\rm RP}-K_s$ to derive its extinction. The dereddened value of
$G_{\rm RP}-K_s$ was used to estimate the spectral type via comparison
to the relation between $G_{\rm RP}-K_s$ and spectral type for members
of Upper Sco \citep{luh18}. Since knowledge of the multiplicity of the members
and candidates is incomplete, $M_J$ was calculated for all stars in a uniform
manner by using seeing-limited photometry from 2MASS and UKIDSS.

\begin{figure}
\epsscale{1.2}
\plotone{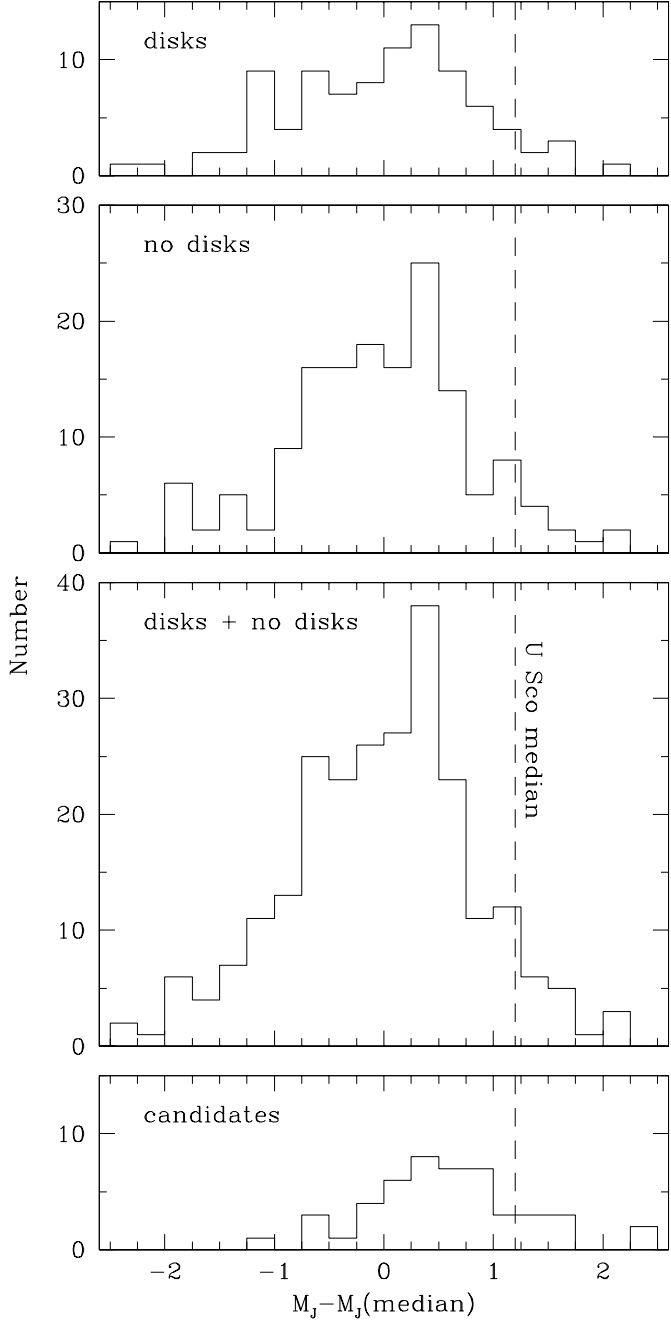}
\caption{
Distributions of offsets of extinction-corrected $M_J$ from the 
median sequence for disk-bearing and diskless members of Taurus
at K0--M7. The distribution for the candidate members above the Tuc-Hor
sequence in Figure~\ref{fig:br} is also shown (Table~\ref{tab:cand}).
The offset of the median sequence for Upper Sco is indicated (dashed line),
which has an age of 11~Myr \citep{pec12}.
}
\label{fig:dj}
\end{figure}

The distributions of $\Delta M_J$ for known members and candidates are shown
in Figure~\ref{fig:dj}. Separate distributions are included for diskless and
disk-bearing members \citep{esp14,esp17}.
One might expect that members with disks would be brighter on average
if they are younger or if disk emission contributes to the observed
fluxes, but the two populations exhibit similar distributions of $\Delta M_J$.
Because of the requirement of a parallax measurement from {\it Gaia}, which
operates at optical wavelengths, the most heavily reddened members are
absent from Figure~\ref{fig:dj}, which tend to be the youngest and least
evolved stars. For instance, most of the $\sim40$ protostars in Taurus lack
parallax measurements. Many of them also lack spectral classifications or
reliable estimates of their extinction-corrected photospheric fluxes.

The $M_J$ offset of the median sequence of Upper Sco \citep[11~Myr,][]{pec12}
relative to Taurus is marked in Figure~\ref{fig:dj}. The value of that
offset is $\sim$1.2~mag, which implies that Taurus is younger by a
factor of $\sim5$ according to evolutionary models of low-mass stars
\citep{sie00,bar98,bar15}.
Very few members of Taurus are fainter than the median of Upper Sco, which
suggests that the number of members with ages of $\gtrsim10$~Myr is quite
small. The true number in that age range may be even smaller than implied
by Figure~\ref{fig:dj} given that some of the faint disk-bearing stars
could have erroneous estimates of their intrinsic fluxes because of scattered
light while a few of the faint diskless stars could be members of 
Cas-Tau that happen to overlap with Taurus in parallax and proper motion
(see Fig.~\ref{fig:map14}).
Measurements of radial velocities for the faintest stars in Figure~\ref{fig:dj} 
would be useful to further constrain their membership.
Meanwhile, the distribution of $\Delta M_J$ for the 
54 candidates from the previous section is somewhat
fainter on average than the known members, indicating older ages.
This difference is a reflection of the fact that a majority of the candidates
are associated with the blue and cyan populations in L1551 and L1517, whose
known members have older median ages than the median of Taurus as a whole
(Figs.~\ref{fig:map13} and \ref{fig:map10}).

The paucity of stars at ages of $\gtrsim10$~Myr in Figure~\ref{fig:dj}
is consistent with previous studies of the distribution of ages in Taurus
\citep{bal99,har01}. The analysis in this work benefits from a larger and
more refined sample of members and better determined completeness at older
ages. Since the stellar populations within Taurus and other molecular clouds
appeared to contain few stars at $\gtrsim10$~Myr and most clouds show
evidence of star formation, \citet{bal99} and \citet{har01b} concluded
that the formation of molecular clouds, the birth of stars within them,
and the dispersal of the clouds all occur rapidly on a timescale of a few Myr.
They found that the small age spread in Taurus was particularly enlightening
since it is much smaller than the crossing time of the region
(few Myr vs. 10--20 Myr), further indicating that molecular clouds form
rapidly, probably through converging flows of atomic gas.
The presence of a small number of older stars is consistent with that
scenario \citep{har12}.

\subsection{Initial Mass Function}
\label{sec:imf}

Previous estimates of the IMF in Taurus have been restricted to specific
areas for which the completeness of the stellar census appeared to be
well-defined \citep{bri02,luh03tau,luh04tau,luh09tau}. For my analysis, I have 
considered the fields in Figures~\ref{fig:map1}--\ref{fig:map15}, which were
searched for new members in Section~\ref{sec:search}. Those fields are
large enough to encompass all of the Taurus clouds and nearly all of the known
members.

\begin{figure}
\epsscale{1.2}
\plotone{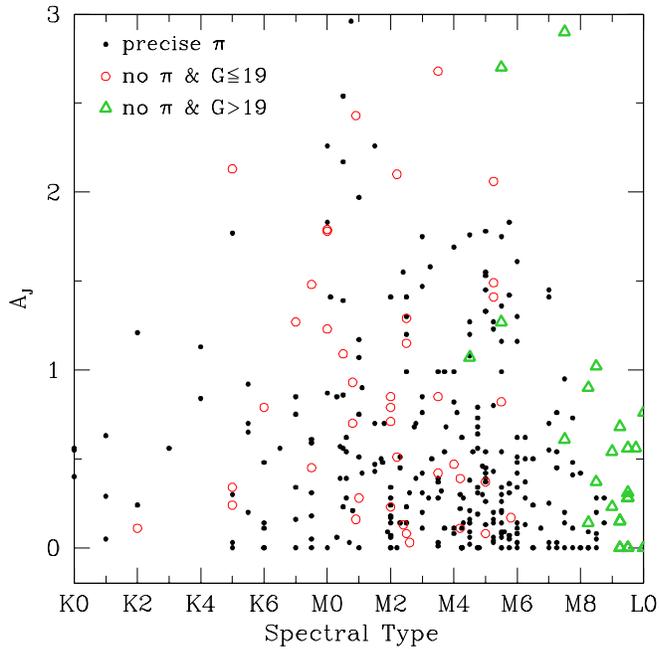}
\caption{
$A_J$ versus spectral type for K0--L0 members of Taurus,
which are labeled according to whether they have precise parallax measurements
from {\it Gaia} DR2 ($\pi/\sigma_\pi\geq10$, filled circles) or lack precise
parallaxes and have $G\leq19$ (red open circles) or $G>19$ (green triangles).
The fraction of stars in DR2 that have parallaxes is high down to $G\sim19$. 
The Taurus members at $G\leq19$ without precise parallaxes have erroneous 
astrometry because of extended emission or close companions.
}
\label{fig:aj}
\end{figure}

A reliable estimate of the IMF in a stellar population requires a sample
of members that is likely to be unbiased in mass.
To identify the selection criteria for such a sample in Taurus,
I examine the completeness of my {\it Gaia} survey for new members.
In {\it Gaia} DR2, most stars have parallax measurements down to $G\sim19$,
and the fraction with parallaxes quickly decreases at fainter magnitudes
\citep{bro18}.
For instance, the percentage of stars in the Taurus fields with parallax
errors of $\leq$0.7~mas ($\lesssim10$\% error at the distance of Taurus)
is $\sim80$\% and $\sim20$\% near $G=19$ and 20, respectively.
The mass (or spectral type) that corresponds to a given limit in $G$ is
a function of extinction, which varies significantly among members of Taurus.
The range of spectral types and extinctions in which Taurus members have
{\it Gaia} parallaxes is illustrated in Figure~\ref{fig:aj}, which shows
extinction versus spectral type for members at K0--L0.  Different symbols
are used for stars with parallaxes that have errors of $\leq$10\% and
that are not discrepant (Section~\ref{sec:pops}) and the remaining members
above and below the magnitude of $G\sim19$ beyond which precise parallaxes
become unavailable. Those three samples contain 289, 44, and 44 stars, 
respectively. Some members cannot be included in Figure~\ref{fig:aj} because
they lack spectral types or extinction estimates, which consist of 
stars with edge-on disks, most protostars, and some companions.
Six of the known members are bright enough at optical wavelengths that they
should be easily detected by {\it Gaia} but do not appear in DR2, 
consisting of HL~Tau, XEST~17-059, J2-2041, V927~Tau, IRAS~04248+2612,
and IRAS~04264+2433. Most of these stars have companions or extended emission
that can account for their absences from DR2. They are excluded from
Figure~\ref{fig:aj}. Similarly, the 44 members in DR2 that have $G\leq19$
but lack parallax measurements have poor astrometric fits, likely due to
companions or extended emission.

\begin{figure}
\epsscale{1.2}
\plotone{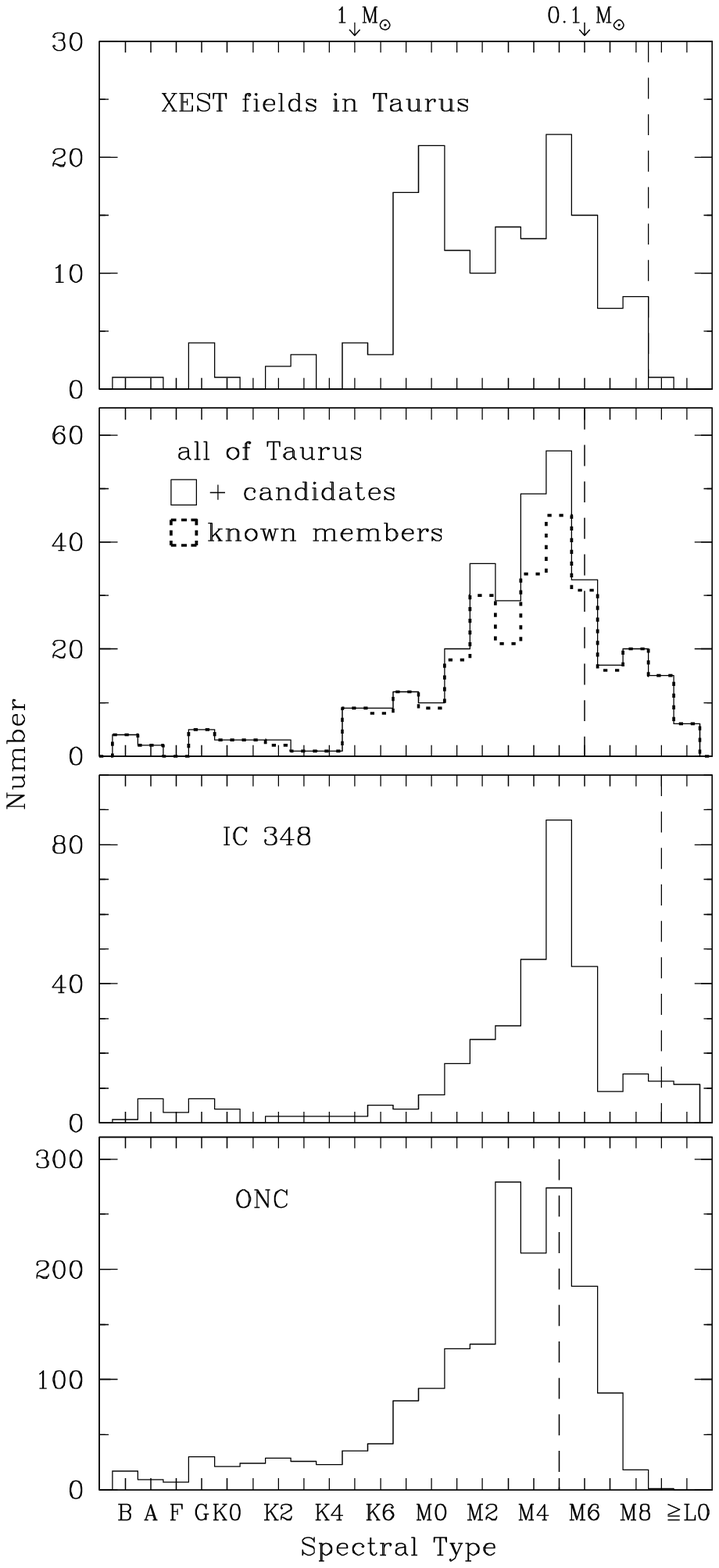}
\caption{
Distributions of spectral types 
for the XEST fields in Taurus as measured by \citet{luh09tau},
the entirety of Taurus for $A_J<1$ (this work),
IC~348 for $A_J<1.5$ \citep{luh16}, and the ONC \citep{dar12,hil13}.
The dashed lines indicate the completeness limits of these samples
and the arrows mark the spectral types that correspond to masses of 0.1
and 1~$M_\odot$ for ages of a few Myr according to evolutionary models
\citep[e.g.,][]{bar98}.
}
\label{fig:imf}
\end{figure}

In Figure~\ref{fig:aj}, the interface between stars with precise parallaxes
and stars at $G>19$ that lack parallaxes (filled circles and open triangles)
roughly approximates the magnitude of $G\sim19$ below which precise
parallaxes become unavailable due to insufficient flux.
That interface extends from $\sim$M9 at $A_J=0$ to mid-M types at $A_J=3$.
I would like to define the IMF sample with an extinction
limit that is high enough to encompass a large number of members but low
enough that the sample has a high level of completeness for {\it Gaia}
parallaxes down to a relatively late spectral type.
Given these considerations, I have selected members with $A_J<1$ for
the IMF sample. As shown in Figure~\ref{fig:aj}, that extinction limit
intersects the $G\sim19$ interface at $\sim$M6--M7, so 
the {\it Gaia} parallaxes (and hence the survey for new members) should be
mostly complete for members within those extinction and spectral type limits.
The completeness limit in spectral type at $A_J=1$ is somewhat uncertain
given the sparse distribution of known late-type members near that extinction.
The more conservative limit of M6 is shown in Figure~\ref{fig:imf}.
If the extinctions of members are independent of mass and spectral type,
then this extinction-limited sample should be unbiased in mass and
representative of the Taurus population.
As mentioned above, some known members at $G\leq19$ lack {\it Gaia} parallaxes,
so the same could be true for undiscovered members. However, the
presence of extended emission is one of the reasons that known members lack
parallaxes, but that is unlikely to be the case for undiscovered members
given that the census of disk-bearing members should be nearly complete
\citep{esp14}. In addition, most of the known members that have erroneous
astrometry due to binarity have spectral types of late-K or early M, whereas
most undiscovered members probably have later spectral types,
which are less prone to binarity-induced astrometric errors due to their
lower binary fractions and smaller separations \citep{duc13}.

For the IMF sample in Taurus, I have selected all known members that
have extinction estimates of $A_J\leq1$ and spectral classifications,
which corresponds to 295 objects.
To avoid some of the sources of uncertainty in estimating masses of young
stars, I use the distribution of spectral types in the sample as a proxy for
the IMF.
In the top panel of Figure~\ref{fig:imf},
I show the distribution reported by \citet{luh09tau} for 26 fields observed
by the {\it XMM-Newton} Extended Survey of
the Taurus Molecular Cloud \citep[XEST,][]{gud07}, which have diameters of
$\sim0.5\arcdeg$ and are centered on the stellar aggregates.
The second panel presents the distribution for my extinction-limited sample
of 295 members across all of Taurus.
That panel also includes the distribution produced after adding the candidate
members with $A_J\leq1$ from Section~\ref{sec:search} (Table~\ref{tab:cand})
using the spectral types and extinctions estimated in Section~\ref{sec:ages}.
For comparison to Taurus, 
I show distributions for an extinction-limited sample in IC~348
\citep{luh16} and a sample of stars in the ONC
in the bottom two panels in Figure~\ref{fig:imf}.
The ONC sample consists of stars from the spectroscopic census 
from \citet{hil13} that are within the $33\arcmin\times33\arcmin$ field
considered by \citet{dar12} and additional stars in that field for which
photometric spectral types were estimated by \citet{dar12}.
Those photometric types should have a high level of completeness down
to $\sim$M5 according to the analysis in \citet{dar12}. 
The field from that study was selected for comparison to Taurus because it
should be large enough to contain a representative sample of members that
is not affected by mass segregation.

As mentioned in Section~\ref{sec:intro} and illustrated in Figure~\ref{fig:imf},
previous samples of Taurus members in small fields like those in the XEST
survey have exhibited a surplus of K7--M0 stars (0.7--0.8~$M_\odot$) 
relative to denser clusters like IC~348 and the ONC.
However, the spectral type distribution for the entirety of Taurus does not
contain a surplus of that kind, and instead resembles the distributions in
IC~348 and the ONC, particularly when the candidates from 
Section~\ref{sec:search} are included. 
Many of the stars with types of K7--M0 in older studies have new
classifications that are later by 1--3~subclasses \citep[][references
therein]{her14,luh17}, which is partially responsible for the disappearance
of the K7--M0 surplus. Mass segregation also appears to be present
in which K7--M0 stars are more likely to be located in smaller areas
surrounding the aggregates than the less massive stars.
Thus, the stellar IMF exhibits little variation between Taurus, IC~348, and
the ONC, which span 3--4 orders of magnitude in stellar density.

\section{Conclusions}

The high-precision astrometry from DR2 of the {\it Gaia} mission has been
used to improve the census of members of the Taurus star-forming region.
The results are summarized as follows.

\begin{enumerate}

\item
Parallaxes and proper motions are available from {\it Gaia} DR2
for 76\% of the stars adopted as Taurus members by \citet{esp17}.
I have used those data to characterize the kinematics and distances
of the groups associated with the Taurus clouds and to check for nonmembers
within that sample. 
After including additional stars that show evidence of membership
from {\it Gaia} and previous spectral classifications,
the revised catalog of members contains 438 objects.

\item
The young stars KPNO~15 and 2MASS J04355209+2255039 have discrepant
proper motions relative to the group in L1536 that they are projected
against. According to their {\it Gaia} proper motions, they
were near the same location within the cloud $\sim7200$ years ago,
indicating that they may have been participants in a dynamical interaction
that resulted in their ejection.

\item
\citet{kra17} and \citet{zha18} presented samples of stars that appear
to be older than the known members of Taurus ($\gtrsim10$~Myr) and that they
classified as members of the region. Among those older stars that have
{\it Gaia} parallaxes and proper motions, none have kinematics and
distances that are consistent with a physical relationship with the
Taurus groups.

\item
A subset of the older stars from \citet{kra17} form a
cluster in parallax and proper motion that coincides with 
a possible new moving group of nine stars found with {\it Gaia} DR1
by \citet{oh17}. I have identified 91 candidate members of this
group using DR2. They have distances of 116--127~pc and an age of
$\sim40$~Myr based on a comparison to Tuc-Hor \citep[45~Myr,][]{bel15}.

\item
I have performed a search for new members of Taurus by selecting stars
from {\it Gaia} DR2 that have proper motions and distances that are
similar to those of any of the groups of known members.
The resulting candidates exhibit two distinct populations, one that
is within the range of ages of the known members ($\lesssim10$~Myr)
and another that is much older ($\gtrsim40$~Myr). 
The latter population is consistent with field stars that are unrelated
to the Taurus clouds.

\item
Relative ages of the known members and candidate members have been
characterized using their offsets in $M_J$ from the median sequence for
Taurus. Very few members or candidates are fainter (older) than the median
sequence of Upper Sco (11~Myr), which contradicts previous reports of a
significant population of older stars ($\gtrsim10$~Myr) associated with
the Taurus groups. The absence of an older population reinforces the
previous evidence that molecular clouds form rapidly on a timescale of a
few Myr \citep{bal99,har01b,har12}.

\item
Previous estimates of the IMF within small fields surrounding the Taurus
aggregates have exhibited a surplus of K7--M0 stars (0.7--0.8~$M_\odot$) 
relative to star-forming clusters like IC~348 and the ONC 
\citep{bri02,luh03tau,luh04tau,luh09tau}.
However, that surplus is absent from the new census for the entire region. 
Thus, the stellar IMF exhibits little variation among nearby star-forming
regions spanning 3--4 orders of magnitude in stellar density.

\end{enumerate}

\acknowledgements
I thank Eric Mamajek for comments on the manuscript.
This work has made use of data from the European Space Agency (ESA)
mission {\it Gaia} (\url{https://www.cosmos.esa.int/gaia}), processed by
the {\it Gaia} Data Processing and Analysis Consortium (DPAC,
\url{https://www.cosmos.esa.int/web/gaia/dpac/consortium}). Funding
for the DPAC has been provided by national institutions, in particular
the institutions participating in the {\it Gaia} Multilateral Agreement.
2MASS is a joint project of the University of Massachusetts and IPAC
at Caltech, funded by NASA and the NSF. 
The Center for Exoplanets and Habitable Worlds is supported by the
Pennsylvania State University, the Eberly College of Science, and the
Pennsylvania Space Grant Consortium.

\clearpage

\begin{deluxetable}{ll}
\tabletypesize{\scriptsize}
\tablewidth{0pt}
\tablecaption{Members of Taurus\label{tab:mem}}
\tablehead{
\colhead{Column Label} &
\colhead{Description}}
\startdata
2MASS & 2MASS Point Source Catalog source name \\
UGCS & UKIDSS Galactic Clusters Survey source name\tablenotemark{a}\\
Names & Other source names \\
RAdeg & Right ascension (J2000) \\
DEdeg & Declination (J2000) \\
Ref-Pos & Reference for right ascension and declination\tablenotemark{b} \\
SpType & Adopted spectral type\tablenotemark{c} \\
pmRA & Proper motion in right ascension from {\it Gaia} DR2\\
e\_pmRA & Error in pmRA \\
pmDec & Proper motion in declination from {\it Gaia} DR2\\
e\_pmDec & Error in pmDec \\
plx & Parallax from {\it Gaia} DR2\\
e\_plx & Error in plx \\
RVel & Radial velocity \\
e\_RVel & Error in RVel \\
r\_RVel & Radial velocity reference\tablenotemark{d} \\
U & $U$ component of space velocity \\
e\_U & Error in U \\
V & $V$ component of space velocity \\
e\_V & Error in V \\
W & $W$ component of space velocity \\
e\_W & Error in W \\
Gmag & $G$ magnitude from {\it Gaia} DR2\\
e\_Gmag & Error in Gmag \\
GBPmag & $G_{\rm BP}$ magnitude from {\it Gaia} DR2\\
e\_GBPmag & Error in GBPmag \\
GRPmag & $G_{\rm RP}$ magnitude from {\it Gaia} DR2\\
e\_GRPmag & Error in GRPmag \\
Pop & Population\tablenotemark{e}
\enddata
\tablenotetext{a}{Based on coordinates from Data Release 10 of the
UKIDSS Galactic Clusters Survey for stars with $K_s>10$ from 
the Two Micron All Sky Survey \citep[2MASS,][]{skr06}.}
\tablenotetext{a}{Sources of the right ascension and declination are 
{\it Gaia} DR2, the 2MASS Point Source Catalog, UKIDSS Data Release 10, and
images from the {\it Spitzer Space Telescope} \citep{luh10tau}.} 
\tablenotetext{c}{Spectral types adopted by \citet{luh17} and \citet{esp17}
with the exception of HD 283641, RX J0422.1+1934, L1551-55, RX J0507.2+2437,
HD 28354, HD 283782, HD 30378,
PSO J065.8871+19.8386/PSO J071.6033+17.0281/PSO J074.1999+29.2197,
and PSO J071.3189+31.6888/PSO J076.2495+31.7503,
whose types are from \citet{wic00}, \citet{mar99}, 
\citet{her14}, \citet{bri99}, \citet{abt04}, 
\citet{wic96}, \citet{rac68}, \citet{esp18}, and this work,
respectively.}
\tablenotetext{d}{
(1) \citet{ngu12};
(2) \citet{gon06};
(3) \citet{muz03};
(4) \citet{whi03};
(5) \citet{tor13};
(6) \citet{har86};
(7) \citet{wic00};
(8) \citet{kra17};
(9) \citet{rei90};
(10) \citet{ric10};
(11) \citet{sce08};
(12) \citet{mat97};
(13) {\it Gaia} DR2.}
\tablenotetext{e}{
Populations in Figures~\ref{fig:map1}--\ref{fig:map15}.}
\tablecomments{
The table is available in a machine-readable form.}
\end{deluxetable}

\begin{deluxetable}{lllcrrrrrrrrr}
\tabletypesize{\scriptsize}
\tablewidth{0pt}
\tablecaption{Median Parallaxes and Proper Motions for Populations in Figures~\ref{fig:map1}--\ref{fig:map11}\label{tab:pmpi}}
\tablehead{
\colhead{Figure} &
\colhead{Pop} &
\colhead{Clouds in Figure} &
\colhead{$\pi$\tablenotemark{a}} &
\colhead{$\mu_\alpha$\tablenotemark{a}} &
\colhead{$\mu_\delta$\tablenotemark{a}} &
\colhead{} &
\colhead{$\Delta\mu_\alpha$\tablenotemark{b}} &
\colhead{$\Delta\mu_\delta$\tablenotemark{b}} &
\colhead{} &
\colhead{$\sigma(\mu_\alpha$)\tablenotemark{c}} &
\colhead{$\sigma(\mu_\delta$)\tablenotemark{c}} &
\colhead{N$_*$}\\
\cline{5-6} \cline{8-9} \cline{11-12} 
\colhead{} &
\colhead{} &
\colhead{} &
\colhead{(mas)} &
\multicolumn{2}{c}{(mas~yr$^{-1}$)} &
\colhead{} &
\multicolumn{2}{c}{(mas~yr$^{-1}$)} &
\colhead{} &
\multicolumn{2}{c}{(mas~yr$^{-1}$/km~s$^{-1}$)} &
\colhead{}}
\startdata
2 & red & B209 & 7.6 &  8.5 & $-$24.4 &  & $-$2.4 & $-$1.0 &  & 1.0/0.6 & 1.5/0.9 & 22 \\
2 & blue & B209 & 6.3 & 12.1 & $-$17.9 &  &  3.0 &  2.1 &  & 0.7/0.5 & 1.2/0.9 & 4 \\
3 & red & L1495 & 7.8 &  8.7 & $-$25.5 &  & $-$1.8 & $-$0.9 &  & 1.5/0.9 & 1.1/0.7 & 31 \\
4 & red & L1521/B213/B215 & 7.6 &  8.0 & $-$23.2 &  & $-$1.5 & $-$0.1 &  & 2.0/1.2 & 2.8/1.7 & 36 \\
4 & blue & L1521/B213/B215 & 6.2 & 11.0 & $-$17.7 &  &  3.1 &  1.5 &  & 4.2/3.2 & 1.7/1.3 & 24 \\
5 & red & L1527 & 7.1 &  5.9 & $-$20.6 &  & $-$0.9 &  0.3 &  & 1.8/1.2 & 2.7/1.8 & 24 \\
6 & red & L1524/L1529/L1536 & 7.8 &  7.2 & $-$21.4 &  & $-$1.0 &  1.5 &  & 1.8/1.1 & 1.2/0.7 & 34 \\
6 & blue & L1524/L1529/L1536 & 6.2 & 10.1 & $-$16.8 &  &  3.6 &  0.8 &  & 1.5/1.1 & 3.1/2.4 & 33 \\
7 & blue & L1489/L1498 & 6.7 & 14.1 & $-$18.8 &  &  3.3 &  0.9 &  & 0.9/0.6 & 0.5/0.4 & 4 \\
8 & red & L1551/L1558 & 6.9 &  6.8 & $-$12.4 &  & $-$2.2 &  5.8 &  & 2.3/1.6 & 1.5/1.0 & 5 \\
8 & blue & L1551/L1558 & 6.9 & 12.0 & $-$18.6 &  &  4.6 & $-$1.6 &  & 1.3/0.9 & 1.3/0.9 & 40 \\
8 & green & L1551/L1558 & 5.1 &  4.8 & $-$13.9 &  &  0.6 & $-$1.3 &  & 0.6/0.6 & 0.3/0.3 & 5 \\
9 & cyan & L1517 & 6.3 &  4.7 & $-$24.5 &  &  0.7 & $-$3.7 &  & 0.8/0.6 & 1.7/1.3 & 32 \\
10 & cyan & L1544 & 5.8 &  2.7 & $-$17.6 &  &  0.0 & $-$0.4 &  & 1.6/1.3 & 2.7/2.2 & 11 
\enddata
\tablenotetext{a}{Based on data from {\it Gaia} DR2.}
\tablenotetext{b}{For a given star, $\Delta\mu$ is defined as the difference
between the {\it Gaia} DR2 proper motion and the proper motion
expected for the position and parallax of the star assuming the median space
velocity of Taurus members from Section~\ref{sec:revised}
($\mu$ $-$ expected $\mu$ in Figures~\ref{fig:map1}--\ref{fig:map11}).}
\tablenotetext{c}{Standard deviation of the proper motions and the
corresponding velocity at the median distance of the population.}
\end{deluxetable}

\begin{deluxetable}{lcccrrrrrrrrrrr}
\tabletypesize{\scriptsize}
\tablewidth{0pt}
\tablecaption{Median Astrometric and Kinematic Parameters for Taurus
Aggregates\label{tab:uvw}}
\tablehead{
\colhead{Population/Cloud} &
\colhead{$\alpha$ (J2000)} &
\colhead{$\delta$ (J2000)} &
\colhead{$\pi$\tablenotemark{a}} &
\colhead{$\mu_\alpha$\tablenotemark{a}} &
\colhead{$\mu_\delta$\tablenotemark{a}} &
\colhead{} &
\colhead{$U$\tablenotemark{b}} &
\colhead{$V$\tablenotemark{b}} &
\colhead{$W$\tablenotemark{b}} &
\colhead{} &
\colhead{$\sigma_U$} &
\colhead{$\sigma_V$} &
\colhead{$\sigma_W$} &
\colhead{N$_*$}\\
\cline{5-6} \cline{8-10} \cline{12-14} 
\colhead{} &
\colhead{(deg)} &
\colhead{(deg)} &
\colhead{(mas)} &
\multicolumn{2}{c}{(mas~yr$^{-1}$)} &
\colhead{} &
\multicolumn{3}{c}{(km~s$^{-1}$)} &
\colhead{} &
\multicolumn{3}{c}{(km~s$^{-1}$)} &
\colhead{}}
\startdata
red/B209 & 63.57 & 28.19 & 7.7 & 8.4 & $-$24.3 &  & $-$15.5 & $-$11.8 & $-$10.6 &  & 2.4 & 0.7 & 1.3 & 6 \\
red/L1495 & 64.73 & 28.40 & 7.8 & 8.7 & $-$25.5 &  & $-$15.8 & $-$12.0 & $-$10.7 &  & 1.9 & 1.2 & 0.9 & 14 \\
red/L1521 & 67.45 & 26.06 & 7.6 & 6.2 & $-$26.0 &  & $-$15.1 & $-$10.5 & $-$9.6 &  & 1.8 & 1.2 & 1.3 & 8 \\
red/L1527 & 69.60 & 25.94 & 7.2 & 5.1 & $-$26.8 &  & $-$15.9 & $-$12.0 & $-$10.5 &  & 2.5 & 1.5 & 0.6 & 4 \\
red/L1524/L1529 & 68.56 & 24.30 & 7.8 & 7.2 & $-$21.2 &  & $-$15.8 & $-$11.1 & $-$9.3 &  & 1.1 & 0.7 & 1.0 & 16 \\
blue/L1536 & 68.96 & 22.84 & 6.2 & 8.4 & $-$21.1 &  & $-$16.8 & $-$13.5 & $-$6.8 &  & 1.1 & 1.3 & 2.0 & 10 \\
blue/L1551 & 68.06 & 18.22 & 6.9 & 12.0 & $-$18.5 &  & $-$15.9 & $-$14.7 & $-$7.6 &  & 1.6 & 1.0 & 0.9 & 18 \\
green/L1558 & 71.75 & 17.00 & 5.1 & 4.8 & $-$20.1 &  & $-$18.9 & $-$13.9 & $-$10.6 &  & 0.6 & 0.3 & 0.3 & 3 \\
cyan/L1517 & 73.94 & 30.37 & 6.3 & 4.3 & $-$24.1 &  & $-$15.0 & $-$14.7 & $-$10.4 &  & 3.3 & 1.2 & 0.9 & 5 
\enddata
\tablenotetext{a}{Based on data from {\it Gaia} DR2.}
\tablenotetext{b}{Based on $UVW$ from Table~\ref{tab:mem}.
The 100 members with estimates of $UVW$ have a median value
of $-15.9, -12.4, -9.4$~km~s$^{-1}$.}
\end{deluxetable}

\begin{deluxetable}{lrrr}
\tabletypesize{\scriptsize}
\tablewidth{0pt}
\tablecaption{Single-star Sequences for the $\beta$~Pic Moving
Group, the Tuc-Hor Association, and the Pleiades Cluster\label{tab:fits}}
\tablehead{
\colhead{$G_{\rm BP}-G_{\rm RP}$} &
\multicolumn{3}{c}{$M_{G_{\rm RP}}$}\\
\cline{2-4}
\colhead{} &
\colhead{$\beta$~Pic} &
\colhead{Tuc-Hor} &
\colhead{Pleiades}}
\startdata
0.0 & 1.40 & 1.40 & 1.40 \\
0.1 & 1.75 & 1.75 & 1.75 \\
0.2 & 2.05 & 2.05 & 2.05 \\
0.3 & 2.25 & 2.25 & 2.25 \\
0.4 & 2.45 & 2.45 & 2.45 \\
0.5 & 2.80 & 2.80 & 2.80 \\
0.6 & 3.15 & 3.15 & 3.15 \\
0.7 & 3.47 & 3.75 & 3.75 \\
0.8 & 3.80 & 4.20 & 4.20 \\
0.9 & 4.05 & 4.60 & 4.60 \\
1.0 & 4.30 & 5.00 & 5.00 \\
1.1 & 4.60 & 5.20 & 5.35 \\
1.2 & 4.90 & 5.34 & 5.65 \\
1.3 & 5.17 & 5.57 & 5.90 \\
1.4 & 5.45 & 5.80 & 6.15 \\
1.5 & 5.65 & 6.00 & 6.40 \\
1.6 & 5.85 & 6.20 & 6.60 \\
1.7 & 6.00 & 6.37 & 6.82 \\
1.8 & 6.20 & 6.54 & 7.05 \\
1.9 & 6.35 & 6.70 & 7.23 \\
2.0 & 6.50 & 6.88 & 7.40 \\
2.1 & 6.65 & 7.04 & 7.55 \\
2.2 & 6.80 & 7.20 & 7.77 \\
2.3 & 7.00 & 7.35 & 7.98 \\
2.4 & 7.20 & 7.55 & 8.15 \\
2.5 & 7.44 & 7.77 & 8.42 \\
2.6 & 7.65 & 8.00 & 8.67 \\
2.7 & 7.85 & 8.30 & 8.90 \\
2.8 & \nodata & 8.60 & 9.20 \\
2.9 & \nodata & 8.85 & 9.55 \\
3.0 & \nodata & 9.15 & 9.90 \\
3.1 & \nodata & 9.45 & \nodata \\
3.2 & \nodata & 9.75 & \nodata \\
3.3 & \nodata & 10.00 & \nodata \\
3.4 & \nodata & 10.30 & \nodata 
\enddata
\tablecomments{The fits are based on data from {\it Gaia} DR2 for members
from \citet{sta07} and \citet[][references therein]{bel15}.
The data for the Pleiades were corrected for extinction assuming
$A_V=0.12$ \citep{sta07} and the reddening relations from \citet{dan18}
and \citet{bab18}. The other two populations should have very little
extinction \citep[$A_V\sim0.03$,][]{bel15}, so corrections were not applied
to their data.}
\end{deluxetable}

\begin{deluxetable}{ll}
\tabletypesize{\scriptsize}
\tablewidth{0pt}
\tablecaption{Candidate Members of Group 29 from \citet{oh17}\label{tab:cand2}}
\tablehead{
\colhead{Column Label} &
\colhead{Description}}
\startdata
2MASS & 2MASS Point Source Catalog source name\\
Name & Other source name \\
RAdeg & Right ascension from {\it Gaia} DR2 (J2000)\\
DEdeg & Declination from {\it Gaia} DR2 (J2000)\\
pmRA & Proper motion in right ascension from {\it Gaia} DR2\\
e\_pmRA & Error in pmRA \\
pmDec & Proper motion in declination from {\it Gaia} DR2\\
e\_pmDec & Error in pmDec \\
plx & Parallax from {\it Gaia} DR2\\
e\_plx & Error in plx \\
Gmag & $G$ magnitude from {\it Gaia} DR2\\
e\_Gmag & Error in Gmag \\
GBPmag & $G_{\rm BP}$ magnitude from {\it Gaia} DR2\\
e\_GBPmag & Error in GBPmag \\
GRPmag & $G_{\rm RP}$ magnitude from {\it Gaia} DR2\\
e\_GRPmag & Error in GRPmag \\
Jmag & $J$ magnitude from the 2MASS Point Source Catalog\\
e\_Jmag & Error in Jmag \\
Hmag & $H$ magnitude from the 2MASS Point Source Catalog\\
e\_Hmag & Error in Hmag \\
Ksmag & $K_s$ magnitude from the 2MASS Point Source Catalog\\
e\_Ksmag & Error in Ksmag 
\enddata
\tablecomments{The table is available in a machine-readable form.}
\end{deluxetable}

\begin{deluxetable}{ll}
\tabletypesize{\scriptsize}
\tablewidth{0pt}
\tablecaption{Candidate Members of Taurus\label{tab:cand}}
\tablehead{
\colhead{Column Label} &
\colhead{Description}}
\startdata
2MASS & 2MASS Point Source Catalog source name\\
RAdeg & Right ascension from {\it Gaia} DR2 (J2000)\\
DEdeg & Declination from {\it Gaia} DR2 (J2000)\\
pmRA & Proper motion in right ascension from {\it Gaia} DR2\\
e\_pmRA & Error in pmRA \\
pmDec & Proper motion in declination from {\it Gaia} DR2\\
e\_pmDec & Error in pmDec \\
plx & Parallax from {\it Gaia} DR2\\
e\_plx & Error in plx \\
Gmag & $G$ magnitude from {\it Gaia} DR2\\
e\_Gmag & Error in Gmag \\
GBPmag & $G_{\rm BP}$ magnitude from {\it Gaia} DR2\\
e\_GBPmag & Error in GBPmag \\
GRPmag & $G_{\rm RP}$ magnitude from {\it Gaia} DR2\\
e\_GRPmag & Error in GRPmag \\
Jmag & $J$ magnitude from the 2MASS Point Source Catalog\\
e\_Jmag & Error in Jmag \\
Hmag & $H$ magnitude from the 2MASS Point Source Catalog\\
e\_Hmag & Error in Hmag \\
Ksmag & $K_s$ magnitude from the 2MASS Point Source Catalog\\
e\_Ksmag & Error in Ksmag 
\enddata
\tablecomments{The table is available in a machine-readable form.}
\end{deluxetable}

\clearpage

\end{document}